\documentclass[graybox]{svmult}

\usepackage{mathptmx}       % selects Times Roman as basic font
\usepackage{helvet}         % selects Helvetica as sans-serif font
\usepackage{courier}        % selects Courier as typewriter font
\usepackage{type1cm}        % activate if the above 3 fonts are
\usepackage[psamsfonts]{amssymb}
\usepackage{amsmath}
\usepackage{amsfonts}
\usepackage{bm}

\usepackage{makeidx}         % allows index generation
\usepackage[dvipdfm]{graphicx}        % standard LaTeX graphics tool
\usepackage{multicol}        % used for the two-column index
\usepackage[bottom]{footmisc}% places footnotes at page bottom
\usepackage{bm}

\makeindex             % used for the subject index
\begin{document}

\title*{Probabilistic flows of inhabitants in urban areas and self-organization in housing markets}
\titlerunning{Self-organization in housing markets}
\author{Takao Hishikawa and Jun-ichi Inoue}
\authorrunning{T. Hishikawa and J. Inoue} 
%for an abbreviated version of
\institute{Takao Hishikawa \at Graduate School of Information Science and Technology, 
Hokkaido University, N14-W-9, Kita-ku, Sapporo 060-0814, Japan 
\and Jun-ichi Inoue \at Graduate School of Information Science and Technology,
Hokkaido University, N14-W-9, Kita-ku, Sapporo 060-0814, Japan \email{jinoue@cb4.so-net.ne.jp}}
\maketitle

\abstract{
We propose a simple probabilistic model to 
explain the spatial structure of the rent distribution of housing market in city of Sapporo. 
Here we modify the mathematical model proposed 
by Gauvin {\it et. al.} \cite{Nadal}.  
Especially, we consider the competition between 
two distances, namely, the distance between house and center, and the distance 
between house and office. 
Computer simulations are carried out to reveal the 
self-organized spatial structure appearing in the rent distribution. 
We also compare 
the resulting distribution with empirical rent distribution in Sapporo as 
an example of cities designated by ordinance.  
We find that the lowest ranking agents (from the viewpoint of the lowest `willing to pay') are 
swept away from relatively attractive regions and make several their own `communities' at  
low offering price locations in the city. 
}
%%%%%%%%%%%%%%%%%%%%%%%%%%%%%%%%%%
%%%%%%%%%%%%%%%%%%%%%%%%%%%%%%%%%%%%%%%%%%%
\section{Introduction}
\label{sec:Intro}
%%%%%%%%%%%%%%%%%%%%%%%%%%%%%%%%%%%%%%%%%%%
Collective behaviour of interacting animals such as flying birds, moving insects or 
swimming fishes has attracted a lot of attentions by scientists and engineers 
due to its  highly non-trivial properties. 
Several remarkable attempts have even done to figure out the mechanism of 
the collective phenomena by collecting empirical data of 
flocking of starlings with extensive data analysis \cite{Ballerini}, 
by computer simulations of realistic flocking  
based on a simple algorithm called BOIDS \cite{Reynolds,Makiguchi}.  
Applications of 
such collective behavior of animals 
also have been proposed in the context of 
engineering \cite{Olfati}. 

Apparently, one of the key factors 
to emerge such non-trivial collective phenomena 
is `local interactions' between agents.  
The local interaction in the microscopic level 
causes non-trivial structures appearing in the macroscopic system. 
In other words, 
there is no outstanding leader who designs the whole system, 
however, 
the spatio-temporal patterns exhibited by the 
system are `self-organized' by local decision making of each 
interacting ingredient in the system. 

These sorts of self-organization by means of local interactions between 
agents might appear not only in natural phenomena 
but also in some social systems including economics. 
For instance, 
decision making of 
inhabitants in urban areas in order to 
look for their houses, the resulting organization of residential street (area) and 
behavior of housing markets are nice examples for 
such collective behavior and emergent phenomena. 
People would search suitable location in their city and decide to 
live a rental (place) if the transaction is approved after negotiation in their own way 
on the rent with buyers. 
As the result, the spatio-temporal patterns 
might be emerged, namely, 
both expensive and cheap residential areas might be co-existed separately in the city. 
Namely, 
local decision makings by 
ingredients --- inhabitants --- 
determine the whole structure of the macroscopic 
properties of the city, 
that is to say, 
the density of residents, 
the spatial distribution of rent, 
and behavior of housing markets. 

Therefore, 
it is very important for us definitely to 
investigate which class of inhabitants chooses 
which kind of locations, rentals, and what is the main 
factor for them to decide their housings. 
The knowledge obtained by answering the above naive questions 
might be useful when we consider the effective urban planning. 
Moreover, 
such a simple but essential question is also important 
and might be an advanced issue in the context of 
the so-called spatial economics \cite{Fujita}. 

In fact, 
constructing 
new landmarks or shopping districts  
might encourage inhabitants to move to a new place to live, 
and at the same time, 
the resulting distribution of residents 
induced by the probabilistic flow of inhabitants who are looking for a new place 
to live might be important information 
for the administrator to consider future urban planning. 
Hence, 
it could be regarded as a typical example of 
`complex systems' 
in which 
macroscopic information (urban planning) and 
microscopic information (flows of inhabitants to look for a new place to live) 
are co-evolved in relatively long time scale under weak interactions.  

To investigate the macroscopic properties of 
the system from the microscopic viewpoint, 
we should investigate the strategy of decision making for individual 
person. 
However, it is still extremely difficult for us to tackle 
the problem by making use of scientifically reliable investigation. 
This is because there exists quite large person-to-person fluctuation 
in the observation of individual behaviour.  
Namely, one cannot overcome the individual variation to 
find the universal fact in the behaviour even though 
several attempts based on impression evaluation or 
questionnaire survey have been done extensively. 
On the other hand, in our human `collective' behaviour instead of individual, 
we sometimes observe several universal facts 
which seem to be suitable materials for computer scientists 
to figure out the phenomena through sophisticated approaches such as agent-based simulations or 
multivariate statistics accompanying with machine learning technique. 

In a mathematical housing market modeling recently proposed by Gauvin {\it et. al.} \cite{Nadal}, 
they utilized several assumptions 
to describe the decision making of each inhabitant in Paris.  
Namely, they assumed that the intrinsic attractiveness of a city 
depends on  
the place and there exists a single peak at the center. 
They also used the assumption that each inhabitant tends to choose the place where the other inhabitants 
having the similar or superior income to himself/herself are living. 
In order to find the best possible place to live, 
each buyer in the system moves from one place to the other according to 
the transition (aggregation) probability described by the above two assumption and 
makes a deal with the seller who presents the best condition for the buyer. 
They concluded that the resulting self-organized rent distribution is almost consistent with 
the corresponding empirical evidence in Paris. 
However, it is hard for us to apply their model directly 
to the other cities having plural centers (not only a single center as in Paris).  

Hence, here we shall modify the Gauvin's model \cite{Nadal} to include 
the much more detail structure of the attractiveness  by taking into account the empirical data 
concerning the housing situation in the city of Sapporo.  
%%%%%%%%%%%%%%%%%%%%%%%%%%%%
\begin{figure}[ht]
\begin{center}
\includegraphics[width=3cm]{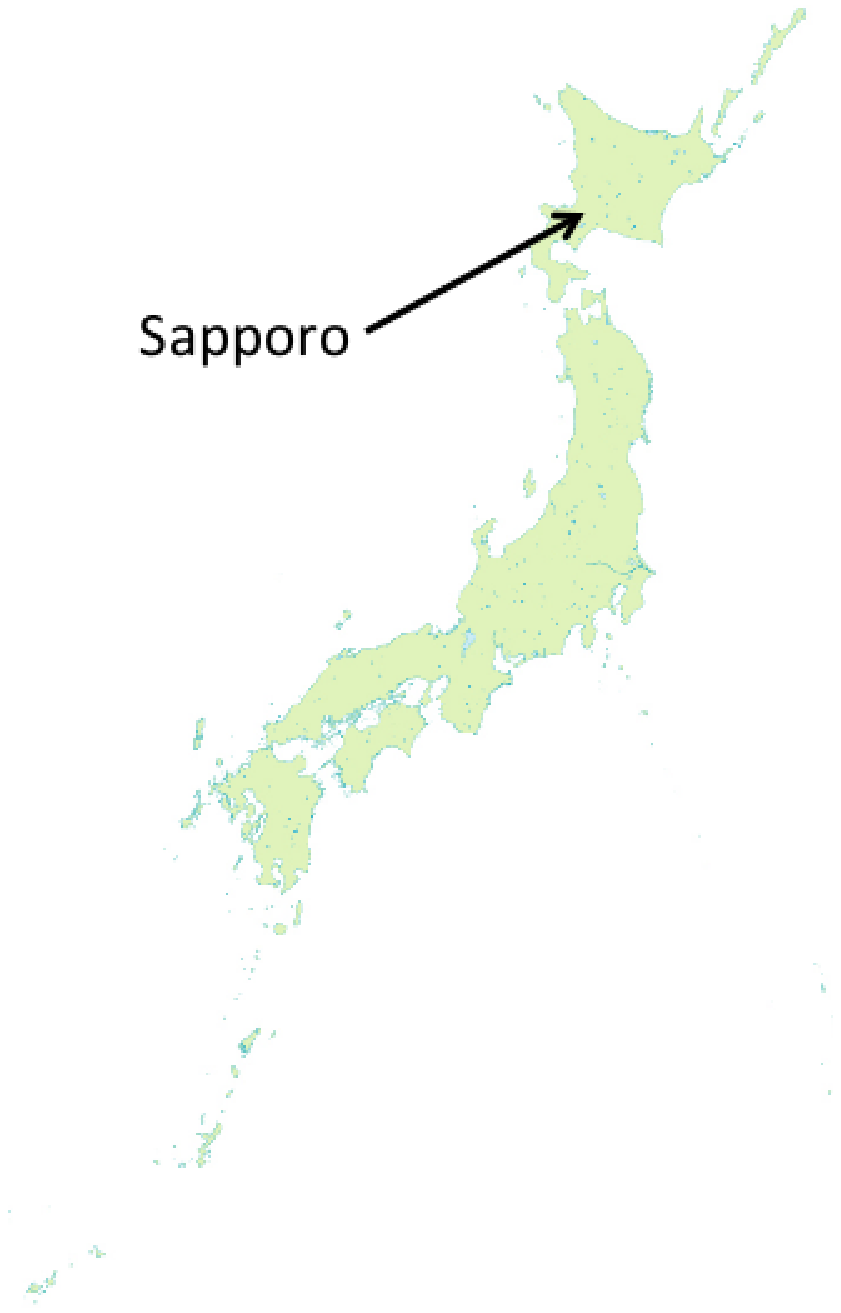}
\includegraphics[width=8.5cm]{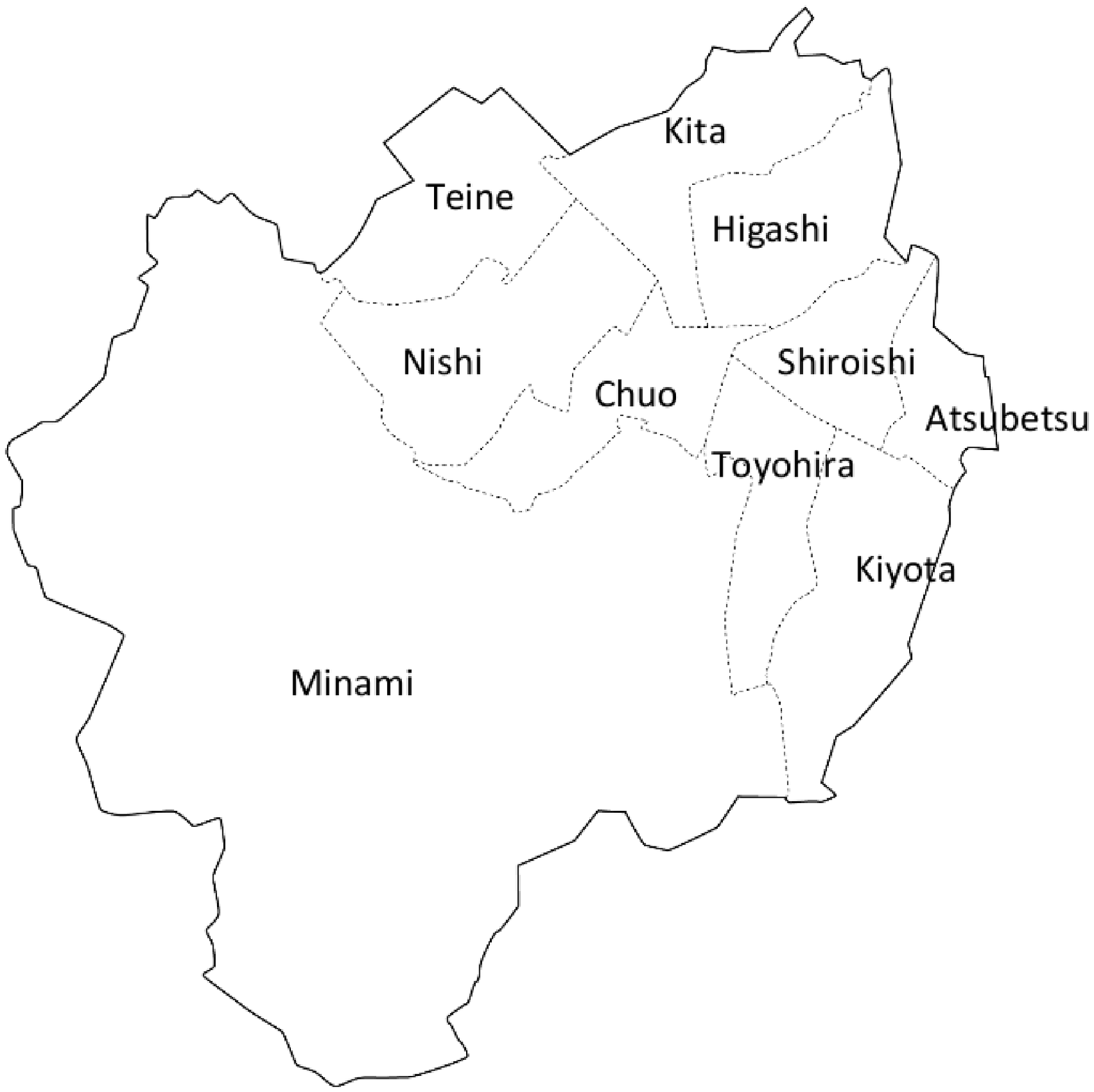}
\end{center}
\caption{\footnotesize 
Sapporo is the largest city on the northern Japanese island of Hokkaido. 
Sapporo is also recognized as 
one of big cities designated by ordinance and 
it has ten wards (we call `ku' for `ward' in Japanese), 
namely, 
{\it Chuo (Central), 
Higashi (East), 
Nishi (West), 
Minami, 
Kita (North), 
Toyohira, 
Shiraishi, 
Atsubetsu, 
Teine} and {\it Kiyota}. 
}
\label{fig:fg_Hokkaido_Sapporo}
\end{figure}
%%%%%%%%%%%
\mbox{}
Sapporo is the fourth-largest city in Japan by population, 
and the largest city on the northern Japanese island of Hokkaido. 
Sapporo is also recognized as 
one of big cities designated by ordinance and 
it has ten wards (we call `ku' for `ward' in Japanese), 
namely, 
{\it Chuo (Central), 
Higashi (East), 
Nishi (West), 
Minami, 
Kita (North), 
Toyohira, 
Shiraishi, 
Atsubetsu, 
Teine} and {\it Kiyota} 
as shown in Fig. \ref{fig:fg_Hokkaido_Sapporo}.

We also consider the competition between 
two distances, namely, the distance between house and center, and the distance 
between house and office. 
Computer simulations are carried out to reveal the 
self-organized structure appearing in the rent distribution. 
Finally, we compare 
the resulting distribution with empirical rent distribution in Sapporo as 
an example of cities designated by ordinance.  
We find that the lowest ranking agents (from the viewpoint of the lowest `willing to pay') are 
swept away from relatively attractive regions and make several their own `communities' at  
low offering price locations in the city. 

This paper is organized as follows. 
In the next section \ref{sec:model}, 
we introduce the Gauvin's model \cite{Nadal} 
and attempt to apply it to explain the 
housing market in city of Sapporo, which is one of 
typical 
cites designated by ordinance in Japan. 
In section \ref{sec:data}, 
we show the empirical distribution of 
averaged rent 
in city of Sapporo and compare the distribution 
with that obtained by computer simulations 
in the previous section \ref{sec:model}. 
In section \ref{sec:extension}, 
we will extend the Gauvin's model \cite{Nadal} 
in which only a single center exists 
to much more generalized model having multiple centers 
located on the places of ward offices. 
In section \ref{sec:simulation}, 
we definitely find that our generalized model 
can explain the qualitative behavior of 
spatial distribution 
of rent in city of Sapporo. 
In the same section, we also show several results 
concerning 
the office locations of inhabitants 
and its effect on the decision making of 
inhabitant moving to a new place. 
The last section 
\ref{sec:summary} is devoted to  summary and discussion. 
%%%%%%%%%%%%%%%%%%%%%%%%%%%%%%%%%%%%%%%%%%%
%%%%%%%%%%%%%%%%%%%%%%%%%%%%%%%%%%
\section{The model system}
\label{sec:model}
%%%%%%%%%%%%%%%%%%%%%%%%%%%%%%%%%%
Here we introduce our model system 
which was originally proposed by 
Gauvin {\it et. al.} \cite{Nadal}. We also 
mention the difficulties we encounter when one applies it to the case of Sapporo city. 
%%%%%%%%%%%%%%%%%%%%%%%%
\subsection{A city --- working space --- }
%%%%%%%%%%%%%%%%%%%%%%%%%%%%%%%
We define our city as a set of nodes 
on the $L \times L$ square lattice.  
The side of each unit of the lattice 
is $1$ and let us call the set as $\Omega$. 
From the definition, the number of elements in the set is given by $|\Omega| \equiv L^{2}$. 
The center of the city is located at $\bm{O} \equiv (L/2,L/2)$. 
The distance between the center 
$\bm{O}$ and the arbitrary place in the city, say, 
$\bm{X} \equiv (x,y)$ is measured by 
%%%%
\begin{equation}
D(\bm{X})=\sqrt{(x-L/2)^{2}+(y-L/2)^{2}}
\end{equation}
%%%%%
where we should keep in mind that 
$D(\bm{X}) \leq L/2$ should be satisfied. 
Therefore, 
if totally $\mathcal{N}$ rentals are on sale in the city, 
 $N \equiv \mathcal{N}/L^{2}$ houses are put up for sale `on average'  
 at an arbitrary  place $\bm{X}$. 
%%%%%%%%%%%%%%%%%%%%%%%%%%%%%%%%%%
 \subsubsection{Why do we choose city of Sapporo?}
%%%%%%%%%%%%%%%%%%%%%%%%%%%%%%%%%%
As we will see later, 
our modeling is applicable to any type of city. 
However, here we choose our home town, 
city of Sapporo, 
as a target city to be examined. 
 %%%%%%
 \begin{figure}[ht]
 \begin{center}
 \includegraphics[width=11cm]{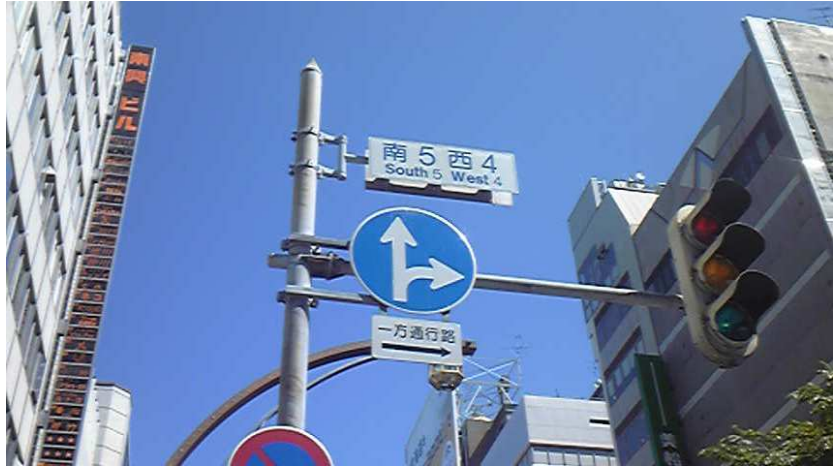}
 \end{center}
 \caption{\footnotesize 
 The mark showing where here is in city of Sapporo (urban district). 
 We can easily find it on the top of traffic signal as 
 `South 5 West 4', 
 which means that here is 
 south by 5 blocks, west by 4 blocks from the origin (`Odori Park', center).
 }
 \label{fig:fg_Sapporo}
 \end{figure}
 %%%%%%%%%%%%%%
 \mbox{}
 
 For the above setting of working space, 
 city of Sapporo is a notable town. 
 This is because the roadways in the urban district are laid to make grid plan road \cite{Sapporo}. 
 Hence, we can easily specify each location by the two-dimensional vector, say, 
 `S5-W4' which means that  south by 5 blocks, west by 4 blocks from the origin (`Odori Park', center). 
 Those labels are usually indicated by marks on the top of the traffic signals (see Fig. \ref{fig:fg_Sapporo}). 
 These kinds of properties might help us to 
 collect the empirical data sets and compare them with 
 the outputs from our probabilistic model. 
%%%%%%%%%%%%%%%%%%%%%%%%%%%%%%%
\subsection{Agents}
\begin{figure}[ht]
\begin{center}
\includegraphics[width=8cm]{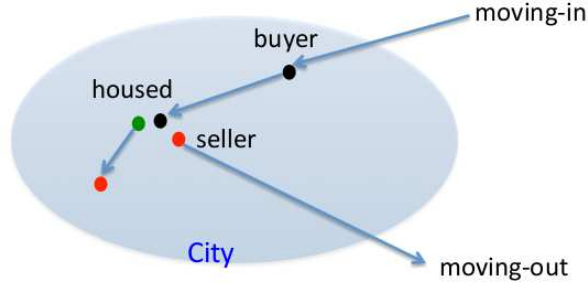}
%\mbox{}\vspace{-0.3cm}
\end{center}
\caption{\footnotesize 
Three kinds of agents in our model system and 
those typical behaviors. 
A newcomer as a `buyer' looks for 
the housing in the city (say, Sapporo). The `housed' who is  
living his/her own house 
becomes a `seller' when 
he/she would move to the other city ({\it e.g.}, Tokyo or Osaka). 
Each seller presents the possible rent to the buyers. 
Once the seller accepts the offer price and makes a contract with the buyer, 
the buyer becomes a `housed' at the place and 
the seller immediately leaves from the city (Sapporo). 
 }
\label{fig:fg_agents}
\end{figure}

%%%%%%%%%%%%%%%%%%%%%%%%%%%%%%
We suppose 
that there co-exist three distinct agents, 
namely, 
`buyers', `sellers' and `housed'. 
The total number of these agents is 
not constant but changing in time. 
For instance, 
in each (unit) time step $t=1$, 
$\Gamma\,\, (\geq 1)$ agents 
visit to the city as `new buyers', 
whereas 
$\alpha$ percentage of the total 
housed persons 
let the house go for some amount of money and 
they become `new sellers'. 
When the `new sellers' succeeded in selling their houses, 
they leave the city to move to the other cities. 
The situation is illustrated by a cartoon in Fig. \ref{fig:fg_agents}. 
%%%%%%%%%%%%%%%%%%
%%%%%%%%%%%%%%%%%%%
\subsubsection{Ranking of agents}
%%%%%%%%%%%%%%%%%%%%%%%%%%%%
Each of agent is categorized (ranked) 
according to 
their degree of `willing to pay' for the housing. 
Let us define the total number of categories 
by $K$. 
Then, 
the agent belonging to the category 
$k \in \{0,\cdots,K-1\}$ can pay 
$P_{k}$ (measured by a currency unit, for instance {\it Japanese yen}) for the housing.  
Hence, we can give the explicit ranking to all agents 
when we put the price $P_{k}$ in the following order 
%%%%
\begin{equation}
P_{0}<P_{1}<\cdots <P_{K-1}.
\end{equation}
%%%%%%
In this paper, 
the price $P_{k}$ of `willing to pay' is given by 
%%%
\begin{equation}
P_{k}=P_{0}+k \, 
\frac{\Delta}{K-1},\,\,\,k=0,\cdots,K-1,  
\end{equation}
%%%%
namely, 
as the difference between 
the highest rent $P_{0}$ 
and the lowest $P_{K-1}$ leads to a gap 
$\Delta$, 
each category (person) is 
put into one of the $\Delta/K$ intervals. 
In our computer simulations which would be given later on, the ranking of each agent is allocated randomly 
from $\{0,\cdots,K-1\}$.  
%%%%
%%%%%%%%%%%%%%%%%%%%%%%%%%%%%%%
 \subsection{Attractiveness of locations}
 %%%%%%%%%%%%%%%%%%%%%%%%%%%%%%%%
 A lot of persons 
 are attracted by 
 specific areas close to 
 the railway (subway) stations 
 or big shopping districts   
 for housing. 
 As well-known, 
 especially in city of Sapporo, 
{\it  Maruyama}-area 
 which is located in the west side of 
 the Sapporo railway 
 station has been appealed to, in particular, 
 high-ranked persons as an exclusive 
 residential district. 
Therefore, one can naturally assume that 
each area in the city possesses 
its own attractiveness and 
we might regard the attractiveness 
as time-independent quantity. 

However, 
the attractiveness might be also dependent on 
the categories (ranking) of agents in some sense. 
For instance, 
the exclusive residential district 
is not attractive for some persons who have relatively lower income and cannot afford the house. 
On the other hand, 
some persons who have relatively higher income 
do not want to live the area where 
is close to the busy street or the slum areas. 

Taking into account these two distinct facts, 
the resulting attractiveness for the area $\bm{X}$ 
should consists of 
the part of the intrinsic attractiveness $A^{0}(\bm{X})$ which is 
independent of the categories $k=0,\cdots,K-1$ and 
the another part of the attractiveness 
which depends on the categories. 
Hence, we 
assume that 
attractiveness at time $t$ for 
the person who belongs to 
the category $k$ at the area $\bm{X}$, that is, 
$A_{k}(\bm{X},t)$ is updated 
by the following spatio-temporal recursion relation: 
% %%%%%
 \begin{equation}
 A_{k}(\bm{X},t+1)=
 A_{k}(\bm{X},t)+
 \omega (A^{0}(\bm{X})-A_{k}(\bm{X},t))+
 \Phi_{k}(\bm{X},t)
 \label{eq:update}
 \end{equation}
%%%%%
where 
$A^{0}(\bm{X})$ 
stands for 
the time-independent intrinsic attractiveness 
for which any person belonging to 
any category 
feels the same amount of charm. 

For example, 
if the center of the city 
$\bm{O}=(L/2, L/2)$
possesses the highest intrinsic 
attractiveness 
$A^{0}_{\rm max}$ as in Paris \cite{Nadal}, 
we might choose 
the attractiveness 
$A^{0}
(\bm{X})$ 
as a two-dimensional Gaussian distribution 
with mean $\bm{O}=(L/2, L/2)$ and 
the variance $R^{2}$ as 
%%%%%
\begin{equation}
A^{0}
(\bm{X})  = 
\frac{A^{0}_{\rm max}}{\sqrt{2\pi R^{2}}}\,
{\exp}
\left[
-\frac{\{
(x-L/2)^{2}+(y-L/2)^{2}\}}
{2R^{2}}
\right]
\label{eq:A0}
\end{equation}
%%%
where the variance 
$R^{2}$ 
denotes a parameter 
which controls the range of influence from 
the center. 
We should notice that 
the equation (\ref{eq:update}) 
has a unique solution 
$A^{0}(\bm{X})$ as a 
steady state 
when we set $\Phi_{k}(\bm{X},t)=0$. 

Incidentally, 
it is very hard for us 
to imagine that 
there exist direct interactions (that is, `communications') between agents who are 
looking for their own houses. 
However,
nobody doubts that 
children's education (schools)
or public peace should be an important issue for 
persons (parents) to look for their  housing. 
In fact, 
some persons 
think that 
their children 
should be brought up 
in a favorable environment 
with their son's/daughter's friends of the same living standard as themselves. 
On the other hand, 
it might be rare 
for persons to move to the area 
where a lot of people who are 
lower living standard than themselves. 
Namely, 
it is naturally assumed 
that people seek for 
the area as their housing place where 
the other people who are higher living standard 
than themselves are living, and if possible, 
they would like to move such area. 

Hence, 
here we introduce such `collective effects'  
into the model system by choosing 
the term $\Phi_{k}(\bm{X},t)$ as 
%%%%%
\begin{equation}
\Phi_{k}(\bm{X},t)= \epsilon 
\sum_{k^{'} \geq k}
v_{k^{'}}(\bm{X},t)
\label{eq:collect}
\end{equation}
%%%%
where $v_{k}(\bm{X},t)$ 
stands for 
the density of housed persons who are in the category $k$ 
at the area $\bm{X}$ 
at time $t$. 
Namely, from equations (\ref{eq:update}) and (\ref{eq:collect}), 
the attractiveness of 
the area $\bm{X}$ for the people of ranking $k$, 
that is, $A_{k}(\bm{X},t)$ 
increases at the next time step $t+1$ 
when persons who are in 
the same as or higher ranking 
than $k$ start to live at $\bm{X}$. 
It should bear in mind 
that 
the intrinsic part of attractiveness 
$A^{0}(\bm{X})$ is time-independent, 
whereas 
$\Phi_{k}(\bm{X},t)$ 
is time-dependent 
through the flows of 
inhabitants in the city. 
Thus, the 
$A_{k}(\bm{X},t)$ could have a different 
shape from the intrinsic part $A^{0}(\bm{X})$ 
due to the effect of the collective behavior of 
inhabitants $\Phi_{k}(\bm{X},t)$. 
%%%%%%%%%%%%%%%%%%%%%%%%%%%
\subsection{Probabilistic search of locations by buyers}
\begin{figure}[ht]
\begin{center}
\includegraphics[width=8cm]{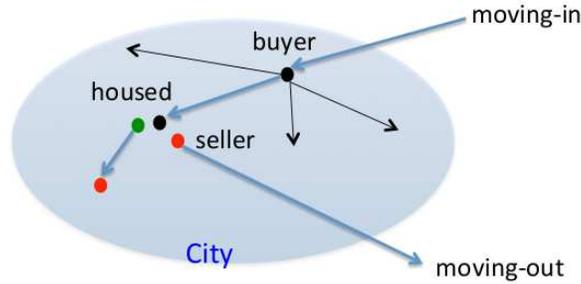}
\end{center}
\caption{\footnotesize 
In each time step, 
a buyer visits 
a place $\bm{X}$ with a probability (\ref{eq:pi}). 
By repeating the transition processes, 
a buyer explores the suitable and desirable location in the city 
from place to place. 
 }
\label{fig:fg_transitions}
\end{figure}

%%%%%%%%%%%%%%%%%%%%%%%%%%%%%%%
The buyers who have not yet determined their own house 
should look for the location. 
Here we assume that 
they move to an arbitrary area 
`stochastically' according to 
the following probability: 
%%%%%%%
\begin{equation}
\pi_{k}(\bm{X},t) = 
\frac{1-{\exp}(-\lambda A_{k}(\bm{X},t))}
{\sum_{\bm{X}^{'} \in \Omega}
\{1-{\exp}(-\lambda A_{k}(\bm{X}^{'},t))\}}
\label{eq:pi}
\end{equation}
%%%
Namely, 
the buyers move to 
the location $\bm{X}$ to look for their housing 
according to the above probability.  
The situation is shown as a cartoon in Fig. 
\ref{fig:fg_transitions}. 
%%%%%%%%%%%%%%%%%%%%%%%
%%%%%%%%%%%%%%
%%%%%% 
We easily find from equation (\ref{eq:pi}) that 
for arbitrary $\lambda$, 
the area which exhibits 
relatively high attractiveness 
is more likely to be selected as a candidate to visit 
at the next round. 
Especially, 
in the limit of $\lambda \to \infty$, 
the buyers who are looking for the locations 
visit only the highest attractiveness 
location 
%%%%
\begin{equation}
\bm{X}_{k}=
{\rm argmax}_{\bm{X}}A_{k}(\bm{X},t). 
\end{equation}
%%%%%%%
On the other hand, 
for $\lambda \ll 1$, 
equation (\ref{eq:pi}) leads to 
%%%%
\begin{equation}
\pi_{k}(\bm{X},t) = 
\frac{A_{k}(\bm{X},t)}
{\sum_{\bm{X}^{'} \in \Omega}
A_{k}(\bm{X}^{'},t)}, 
\end{equation}
%%%%
and the probability for buyers to visit the place $\bm{X}$ at time $t$ is proportional to 
the attractiveness corresponding to 
the same area $A_{k}(\bm{X},t)$.  

Here we should mention that 
when we regard the `location' as `company', 
the present model system is described by similar aggregation probability 
to the probabilistic labor market 
proposed by Chen {\it et. al.} \cite{Chen} 
where the parameter $\gamma$ corresponds to 
the $\lambda$ in the Gauvin's model \cite{Nadal}. 
%%%%%%%%%%%%%%%%%%%%%
\subsection{Offering prices by sellers}
%%%%%%%%%%%%%%%%%%%%%%%%%%
It is not always possible for buyers to  
live at the location $\bm{X}$ where 
they have selected to visit according to 
the probability 
$\pi_{k}(\bm{X},t)$ 
because 
it is not clear whether they can accept the 
price offered by the sellers at $\bm{X}$ or not. 
Obviously, 
the offering price itself 
depends on 
the ranking $k$ of the sellers. 
Hence, here we assume that 
the sellers of 
ranking $k$ at the location $\bm{X}$ 
offer the buyers the price: 
%%%%%
\begin{equation}
P_{k}^{o} (\bm{X}) = 
P^{0} + 
[1-{\exp}(-\xi \overline{A}(\bm{X},t))]P_{k}
\end{equation}
%%%%
for the rental housing, 
where 
$\overline{A}(\bm{X},t)$ means 
the average of the attractiveness 
$A_{k}(\bm{X},t)$ over the all categories 
$k=0,\cdots, K-1$, 
that is to say, 
%%%
\begin{equation}
\overline{A}(\bm{X},t) \equiv 
\frac{1}{K}\sum_{k=0}^{K-1}
A_{k}(\bm{X},t), 
\end{equation}
%%%%
and $\xi$ is a control parameter. 
In the limit of 
$\xi \to \infty$, 
the offering price 
by a seller of ranking $k$ 
is given by 
the sum of 
the basic rent $P^{0}$ and 
the price of `willing to pay' for 
the sellers of ranking $k$, namely, 
$P_{k}$ as $P_{k}^{o}=P^{0}+P_{k}$. 

For the location $\bm{X}$
in which any transaction 
has not yet been approved, 
the offering price is 
given as 
%%%%%
\begin{equation}
P_{k}^{o} (\bm{X}) = 
P^{0} + 
[1-{\exp}(-\xi \overline{A}(\bm{X},t))]P^{1}
\end{equation}
%%%
because 
the ranking of the sellers is ambiguous 
for such location $\bm{X}$. 
%%%%%%%%%%%%%%%%%%%%%%%%%%%%%%%%
\subsection{The condition on which the transaction is approved}
%%%%%%%%%%%%%%%%%%%%%%%%%%%%%
It is needed for buyers to 
accept the price offered by sellers in order to 
approve the transaction. 
However, if the offering price is higher 
than the price of `willing to pay' for the buyer, 
the buyer cannot accept the offer. 
Taking into account this limitation, 
we assume that 
the following condition 
between 
the buyer of the ranking $k$ and the seller 
of the ranking $k^{'}$ 
at the location $\bm{X}$ 
should be 
satisfied to approve the transaction. 
%%%%%
\begin{equation}
P_{k} > P_{k^{'}}^{o}(\bm{X})
\label{eq:condition}
\end{equation}
%%%%
Thus, if and only if the above condition 
(\ref{eq:condition}) 
is satisfied, 
the buyer can own the housing 
(the seller can sell the housing). 

We should keep in mind that 
there is a possibility for the lowest ranking 
people to fail to 
own any housing in the city 
even if they negotiate with the person 
who also belongs to the lowest ranking. 
To consider the case 
more carefully, let us set $k=k^{'}=0$ in the 
condition (\ref{eq:condition}), 
and then we have 
$P_{0} > {P^{0}}/{{\exp}[-\xi \overline{A}(\bm{X},t)]}$. 
Hence, 
for the price of `willing to pay' 
$P_{0}$ of the lowest ranking people, 
we should determine the lowest rent  
$P^{0}$ so as to satisfy  
$P^{0} < {\exp}[-\xi \overline{A}(\tilde{\bm{X}},t)]P_{0}$. 
Then, the lowest ranking people never fail to 
live in the city. 

We next  define the actual transaction 
price as interior division point between 
$P_{k}$ and $P_{k^{'}}^{o}(\bm{X})$ 
by using a single parameter $\beta$ as 
%%%%
\begin{equation}
P_{\rm tr}=(1-\beta)P_{k^{'}}(\bm{X}) + 
\beta P_{k}. 
\label{eq:Ptr}
\end{equation}
%%%
By repeating these 
three steps, 
namely, 
probabilistic searching 
of the location by buyers, 
offering the rent by sellers, 
transaction between 
buyers and sellers 
for enough times, 
we evaluate the 
average rent at each 
location 
$\bm{X}$ and 
draw 
the density of inhabitants, 
the spatial distribution of 
the average rent in two-dimension. 
In following, 
we show our preliminary results. 
%%%%%%%%%%%%%%%%%%%%%%%%%%%%%
\subsection{Computer simulations: A preliminary}
\begin{figure}[ht]
\begin{center}
\includegraphics[width=10cm]{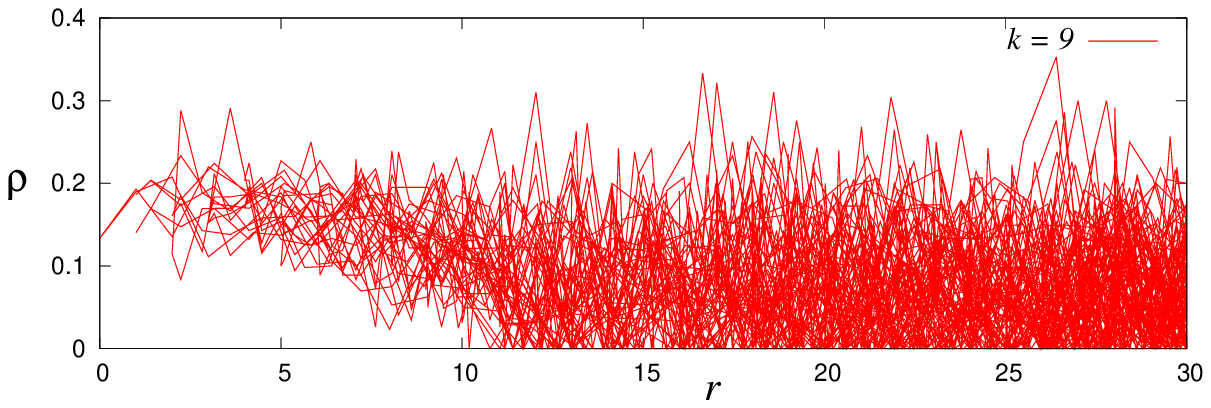} 
%\mbox{}\vspace{-0.7cm} \\
\includegraphics[width=10cm]{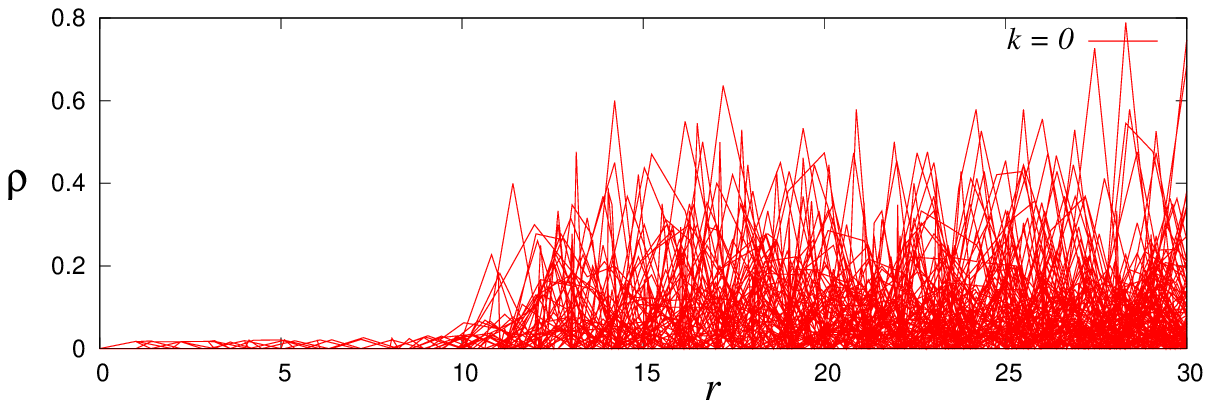}
\end{center}
%\mbox{}\vspace{-0.7cm}
\caption{\footnotesize The spatial density $\rho(r)$ of 
inhabitants  obtained by the Gauvin's model \cite{Nadal} 
having a single center in the city. 
The horizontal axis 
$r$ stands for 
the distance between the center $\bm{O}=(L/2,L/2)$ and 
the location $\bm{X}=(x,y)$, namely, 
$r=D(\bm{X})$. 
The upper panel 
is the result for the highest raining people ($k=K-1$), 
whereas the lower panel shows 
the result for the lowest ranking inhabitants ($k=0$). 
We easily recognize that 
the persons who are belonging to the lowest ranking 
cannot live the area close to the center. 
}
\label{fig:fg0}
%\mbox{}\vspace{-0.7cm}
\end{figure}

%%%%%%%%%%%%%%%%
In Fig. \ref{fig:fg0}, 
we show the spatial density distribution $\rho (r)$ of 
inhabitants by the Gauvin's model \cite{Nadal} 
having a single center in the city. 
The horizontal axis $r$ of these panels 
stands for 
the distance between the center $\bm{O}$ and 
the location $\bm{X}$, namely, 
$r=D(\bm{X})$. 
We set the parameters appearing in the model 
as 
$L=100, \alpha =0.1, 
K=10, 
P_{0}=15,000, 
\Delta +P_{0}= 120,000 \,(\Delta=105,000), 
\omega = 1/15, 
R=10, \epsilon=0.0022, 
\lambda=0.01, 
\zeta=0.1, P^{0}=9,000,
P^{1}=200,000, 
\beta=0.1, \Gamma=L^{2}/K, \\
T=100\, 
(\equiv \mbox{Total number of updates (\ref{eq:update})})$.   
It should be noted that the definition of density 
is given by 
%%%%
\begin{equation}
\rho (r) \equiv 
\frac{\mbox{($\#$ of inhabitants of 
ranking $k$ at the location $r$)}}{
\mbox{(Total $\#$ of inhabitants at the location $r$)}}.
\end{equation}
%%%
%%%%%%
%%%%%
From the lower panel, 
we easily recognize that 
the persons who are belonging to the lowest ranking 
cannot live the area close to the center in the city. 
Thus, we find that there exists a clear division of 
inhabitants of different rankings. 

Intuitively, these phenomena 
might be understood as follows. 
The persons of the highest ranking ($k=K-1$) 
can afford to accept any offering price 
at any location. 
At the same time, 
the effect of aggregation induced by $\Phi_{K-1}(\bm{X},t)$ 
in the searching probability $\pi_{K-1}(\bm{X},t)$ and 
the update rule of $A_{K-1}(\bm{X},t)$ in (\ref{eq:update}) is 
the weakest among the $K$ categories. 
As the result, 
the steady state of 
the update rule (\ref{eq:update}) is 
not so deviated 
from the intrinsic attractiveness, 
namely, $A_{K-1}(\mbox{\boldmath $X$},t) \simeq A^{0}(\bm{X})$. 
Hence,  people of 
the highest ranking 
are more likely to visit the locations where are close to the center 
$\bm{X}=\bm{O}$ and they live there. 
Of course, 
the density of the highest ranking persons 
decreases as $r$ increases. 

On the other hand, people of the lowest ranking ($k=0$) 
frequently visit the locations  
where various ranking people are living due to 
the aggregation effect of the term 
$\Phi_{0}(\bm{X},t)$ to look for their housing.  
However, 
it is strongly dependent on 
the offering price at the location 
whether the persons can live there or not. 
Namely, the successful rate 
of transaction depends on 
which ranking buyer offers the price 
at the place $\bm{X}$.  
For the case in 
which 
the `housed' people changed to sellers at the location where is 
close to the center, 
the seller is more likely to be the highest ranking person 
because he/she was originally an inhabitant 
owing the house near the center. 
As the result, 
the price offered by them 
might be too high for the people of the lowest ranking 
to pay for approving the transaction. 
Therefore, 
the lowest raining people might be driven away to 
the location where is far from the center.  
%%%%%%%%%%%%%%%%%%%%%%%%%%%%%%%%%%
\section{Empirical data in city of Sapporo}
\label{sec:data}
\begin{figure}[ht]
\begin{center}
\includegraphics[width=9cm]{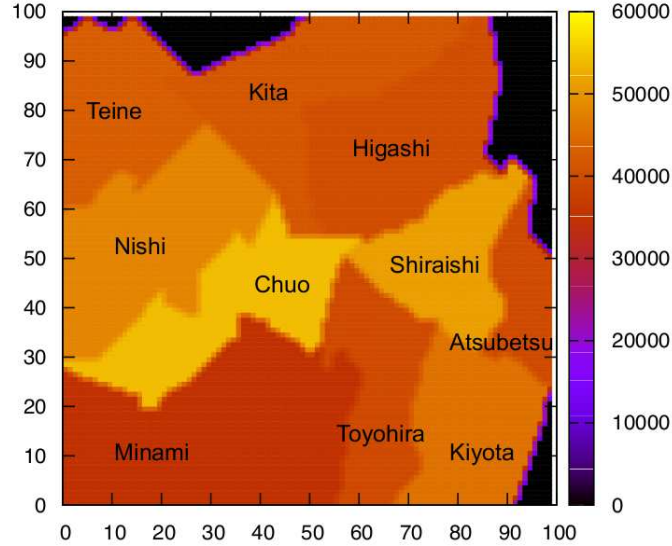}
\end{center}
%\mbox{}\vspace{-0.7cm}
\caption{\footnotesize 
The spatial distribution of 
the averaged rent in city of Sapporo  
by using 
the empirical housing data collected from \cite{Souba}. 
In Sapporo, 
ten wards 
{\it Chuo (Central), 
Higashi (East), 
Nishi (West), 
Minami (South), 
Kita, 
Toyohira, 
Shiraishi, 
Atsubetsu, 
Teine} and {\it Kiyota} 
exist.  (COLOR ONLINE)
}
%\mbox{}\vspace{-0.7cm}
\label{fig:fg02}
\end{figure}

%%%%%%%%%%%%%%%%%%%%%%%%%%%%%%%%%%
Apparently, 
the above simple modeling with a single 
center of city is limited to the specific class of city like Paris. 
Turning now to the situation in Japan, there are several major cites 
designated by ordinance, and 
city of Sapporo is one of such `mega cities'. 
In Table \ref{tab:population}, 
we show several statistics in Sapporo in 2010. 
From this table, we recognize that 
in each year, 
63,021 persons are moving into and 
57,587 persons are moving out from Sapporo. 
Hence, the population in Sapporo is still increasing by approximately six thousand in each year. 
%%%%%%%%%%%%%%%%%%%%%%%%%%%%%%%%%
\begin{table}[htbp]
\begin{center}
\begin{tabular}{|l|r|r||r|r|} \hline
  Wards & $\#$ of moving-in & $\#$ of moving-out & lowest (yen) & highest  (yen) \\ \hline 
    {\it Chuo (Central)} & 12,132 & 10,336 & 19,000 & 120,000 \\ \hline
    {\it Kita (North)} & 8,290 & 7,970 & 15,000 & 73,000 \\ \hline
    {\it Higashi (East)}  & 7,768 & 7,218 & 20,000 & 78,500 \\ \hline
    {\it Shiraishi} & 6,857 & 6,239 & 25,000 & 67,000 \\ \hline
    {\it Atsubetsu} & 4,003 & 3,736 & 33,000 & 57,000 \\ \hline
    {\it Toyohira} & 7,854 & 7,037 & 20,000 & 69,000 \\ \hline
    {\it Kiyota} & 2,560 & 2,398 & 30,000 & 55,000 \\ \hline 
    {\it Minami (South)}  & 3,824 & 3,794 & 23,000 & 58,000 \\ \hline
    {\it Nishi} & 6,315 & 5,788 & 20,000 & 80,000 \\ \hline
    {\it Teine} & 3,418 & 3,071 & 20,000 & 68,000 \\ \hline \hline
    Total  & 63,021 & 57,587 & --- & --- \\ \hline
\end{tabular}
\end{center}
\caption{\footnotesize 
Statistics in city of Sapporo for the number of persons who were moving-into and -out, 
the lowest and highest rents (the unit is Japanese yen) of 2DK-type flats in city of Sapporo.  
}
\label{tab:population}
\end{table}
%%%%
As we already mentioned, Sapporo is the fourth-largest city in Japan by population, 
and the largest city on the northern Japanese island of Hokkaido. 
Sapporo is recognized as 
one of big cities designated by ordinance and 
it has ten wards (what we call `ku' in Japanese), 
namely, 
{\it Chuo, 
Higashi, 
Nishi, 
Minami, 
Kita, 
Toyohira, 
Shiraishi, 
Atsubetsu, 
Teine} and {\it Kiyota} 
as shown in Table \ref{tab:population} 
(for details, see \cite{Sapporo} for example). 
Hokkaido prefectural office is 
located in {\it Chuo}-ku and 
the other important landmarks 
concentrate in the wards. 
Moreover, as it is shown in Table \ref{tab:population}, the highest and 
the lowest rents for the 2DK-type (namely,  a two-room apartment with a kitchen/dining area) flats in {\it Chuo}-ku are both the highest among ten wards. 
In this sense, {\it Chuo}-ku could be regarded as 
a `center' of Sapporo. 
However, the geographical structure of 
rent distribution in city of Sapporo is far from 
the symmetric one as given by the intrinsic attractiveness  $A^{0}(\bm{X})$ 
having a single center (see equation (\ref{eq:A0})). 
In fact, we show the rough distribution of average rents in city of Sapporo in Fig.\ref{fig:fg02} 
by making use of the empirical data collected from \cite{Souba}. 
From this figure, we clearly confirm that 
the spatial structure of rents in city of Sapporo is not symmetric but apparently asymmetric. 
%%%%%
%%%%%%%%%%%%%%%%%

From this distribution, 
we also find that 
the average rent is dependent on 
wards,  and actually it is very hard to simulate 
the similar spatial distribution 
by using the Gauvin's model \cite{Nadal} 
in which there exists only a single center in the city. 
This is because 
in a city designated by ordinance like Sapporo, 
each ward formulates its own 
community, 
and in this sense, each ward should be regarded as 
a `small city' having a single (or multiple) center(s). 
It might be one of the essential differences between Paris and Sapporo. 
%%%%%%%%%%%%%%%%%%%%%%
%%%%%%%%%%%%%%%%%%%%%%%%%%%%%%%%%
\section{An extension to a city having multiple centers}
\label{sec:extension}
%%%%%%%%%%%%%%%%%%%%%%%%%%%%%%%%%
In the previous section \ref{sec:data}, 
we found that the Gauvin's model \cite{Nadal} 
having only a single center 
is not suitable to explain 
the empirical evidence for 
a city designated by ordinance such as 
Sapporo 
where multiple centers as wards co-exist.  

Therefore, 
in this section, we modify the intrinsic attractiveness $A^{0}(\bm{X})$ 
to explain the empirical evidence in city of Sapporo. 
For this purpose, 
we use the label $l=1,\cdots,10$ 
to distinguish ten words in Sapporo, namely, 
 {\it Chuo (Central), 
Higashi (East), 
Nishi (West), 
Minami (South), 
Kita (North), 
Toyohira, 
Shiraishi, 
Atsubetsu, 
Teine} and {\it Kiyota} 
in this order, and define 
$\bm{B}_{l}=(x_{B_{l}},y_{B_{l}}),l=1,\cdots, 10$ 
for each location
where each ward office is located. 
Then, we shall modify 
the intrinsic attractiveness in terms of 
the $\bm{B}_{l}$ as follows. 
%%%%%
\begin{equation}
A^{0}(\bm{X}) = 
\sum_{l=1}^{10}
\frac{\delta_{l}}{\sqrt{2\pi} R_{l}}
{\exp}
\left[
-\frac{
\{
(x-x_{B_{l}})^{2}+
(y-y_{B_{l}})^{2}
\}
}{2R_{l}^{2}}
\right]
\label{eq:ours}
\end{equation}
%%%%%
where 
%%%
%%%%
\begin{equation}
\delta_{1}+\cdots + \delta_{10}=1
\label{eq:normal2}
\end{equation}
%%%
should be satisfied. 
Namely, 
we would represent the intrinsic attractiveness $A^{0}(\bm{X})$ 
in city of Sapporo by means of 
a two-dimensional mixture of Gaussians 
in which each mean corresponds to 
the location of the ward office. 
$R_{l}\, (l=1,\cdots,10)$
denotes a set of parameters 
which control the spread of the center. 
In our computer simulations, 
we set $R_{l}=5$ for all 
$l=1,\cdots,10$ in our intrinsic attractiveness (\ref{eq:ours}). 

Here we encounter a problem, 
namely, we should choose each weight 
$\delta_{l},\,l=1,\cdots,10$. 
For this purpose, we see the number of 
estates (flats) in each wards. 
From a real-estate agents in Sapporo
\cite{Homes}, we have the statistics as {\it Chuo (Central)} ($9,598$), 
{\it Higashi (East)} ($6,433$), 
{\it Nishi (West)} ($5,830$), 
{\it Minami (South)} ($1,634$), 
{\it Kita (North)} ($4,671$), 
{\it Toyohira} ($4,893$), 
{\it Shiraishi} ($5,335$), 
{\it Atsubetsu} ($1,104$), 
{\it Teine} ($2,094$), 
{\it Kiyota} ($962$). 
Hence, by dividing each number by the maximum $9,598$ for {\it Chuo}-ku, 
the weights  $\delta_{l},\,l=1,\cdots, 10$ are chosen as  
{\it Chuo (Central)} ($\delta_{1} \propto 1.0$), 
{\it Higashi (East)} ($\delta_{2} \propto 0.67$), 
{\it Nishi (West)} ($\delta_{3} \propto 0.61$), 
{\it Minami (South)} ($\delta_{4} \propto 0.17$), 
{\it Kita (North)} ($\delta_{5} \propto 0.49$), 
{\it Toyohira} ($\delta_{6} \propto 0.51$), 
{\it Shiraishi} ($\delta_{7} \propto 0.56$), 
{\it Atsubetsu} ($\delta_{8} \propto 0.12$), 
{\it Teine} ($\delta_{9} \propto 0.22$), 
{\it Kiyota} ($\delta_{10} \propto 0.1$). 
Of course, we normalize these parameters 
so as to satisfy the condition (\ref{eq:normal2}). 
%%%%%%%%%%%%%%%%%
\section{Computer simulations}
\label{sec:simulation}
%%%%%%%%%%%%%%%%%%%%%%%%
In this section, we show the results of computer simulations. 
%%%%%%%%%%%%%%%%%%%%%%%%%%%
\subsection{Spatial structure in the distribution of visiting times}
\begin{figure}[ht]
\begin{center}
%%%%%
 \includegraphics[width=4.7cm]{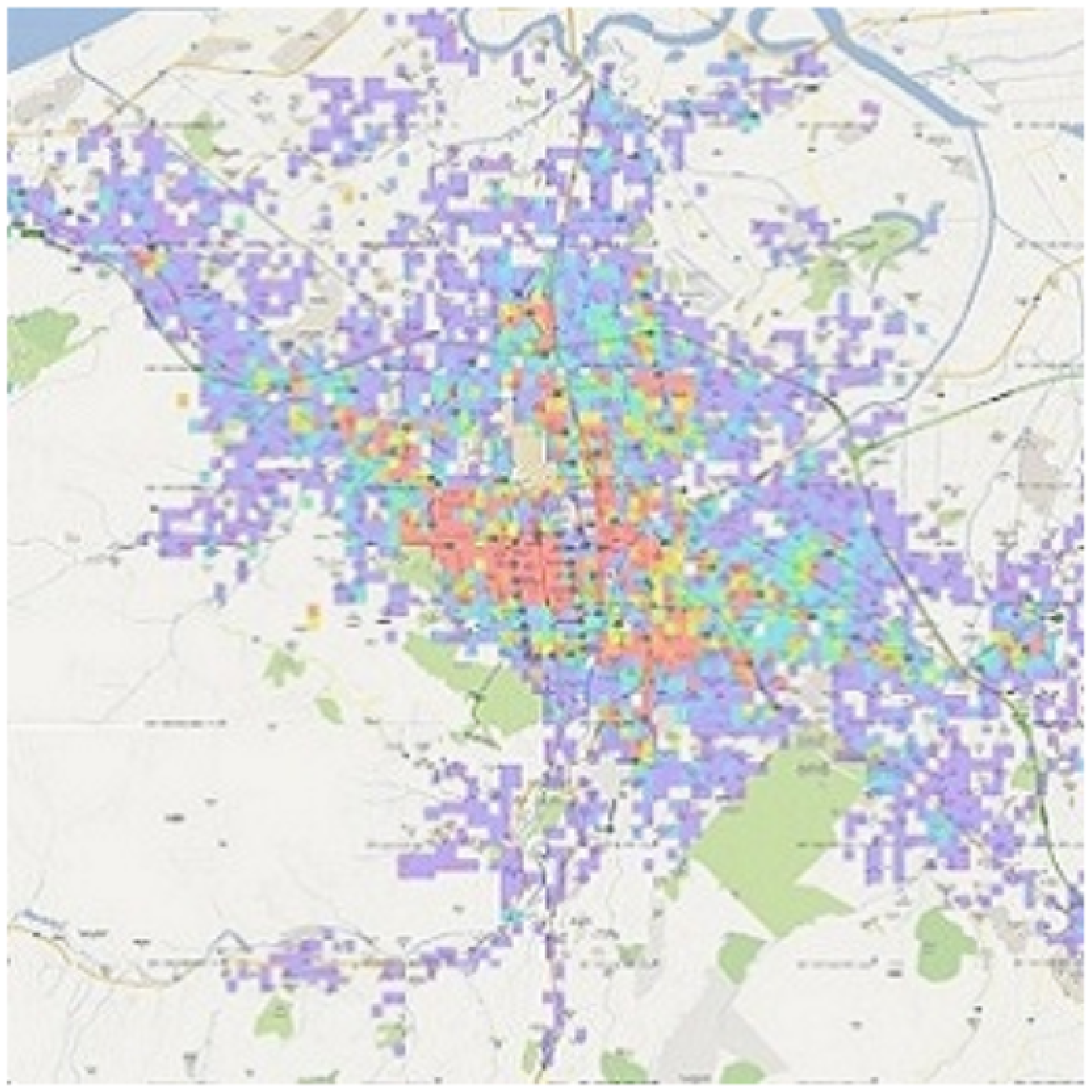} 
 %\hspace{-0.7cm}
\includegraphics[width=5.65cm]{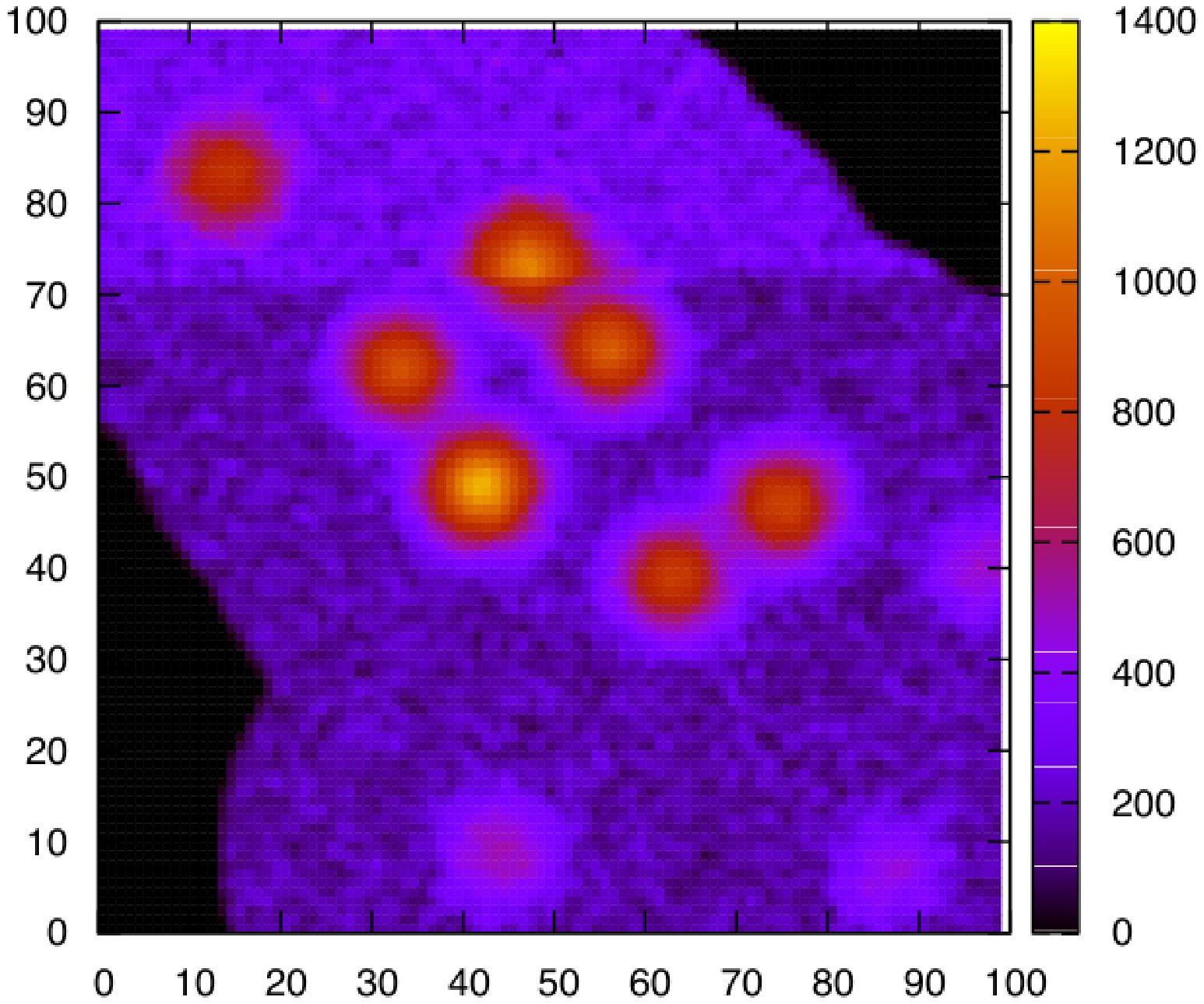}
%\mbox{}\vspace{-0.7cm}
\end{center}
\caption{\footnotesize  
The distributions of the number of persons who checked the information about the flat located at $\bm{X}$  on the web site (left, from \cite{Homes}) 
and the number of persons who visited the place $\bm{X}$ according to 
the transition probability (\ref{eq:pi}) (right) in our artificial society. (COLOR ONLINE)}
\label{fig:DemandMap}
%\end{minipage}
%\mbox{}\vspace{-0.5cm}
\end{figure}

%%%%%%%%%%%%%%%%%%%%%%%%%%%%%%%%%%%
In Fig. \ref{fig:DemandMap} (left), 
we show the distributions of the number of persons 
who checked the information about the flat located at $\bm{X}$ 
and the number of persons who visited the place $\bm{X}$ according to 
the transition probability (\ref{eq:pi}) in the right panel. 
From this figure, 
we find that 
the locations (flats)  $\bm{X}$ where people 
checked on the web site \cite{Homes} most frequently concentrate 
around each ward office. 
From this fact, our modeling 
in which we choose the locations of 
multiple centers as the places of wards might be approved. 
Actually, 
from the right panel of 
this figure, we are confirmed that 
the structure of 
spatial distribution is qualitatively very similar to 
the counter-empirical distribution (right panel). 
%%%%%
%%%%%
%%%%%%%%%%%%%%%%%
\mbox{}

\begin{figure}[h]
\begin{center}
%%%
\includegraphics[width=5.6cm]{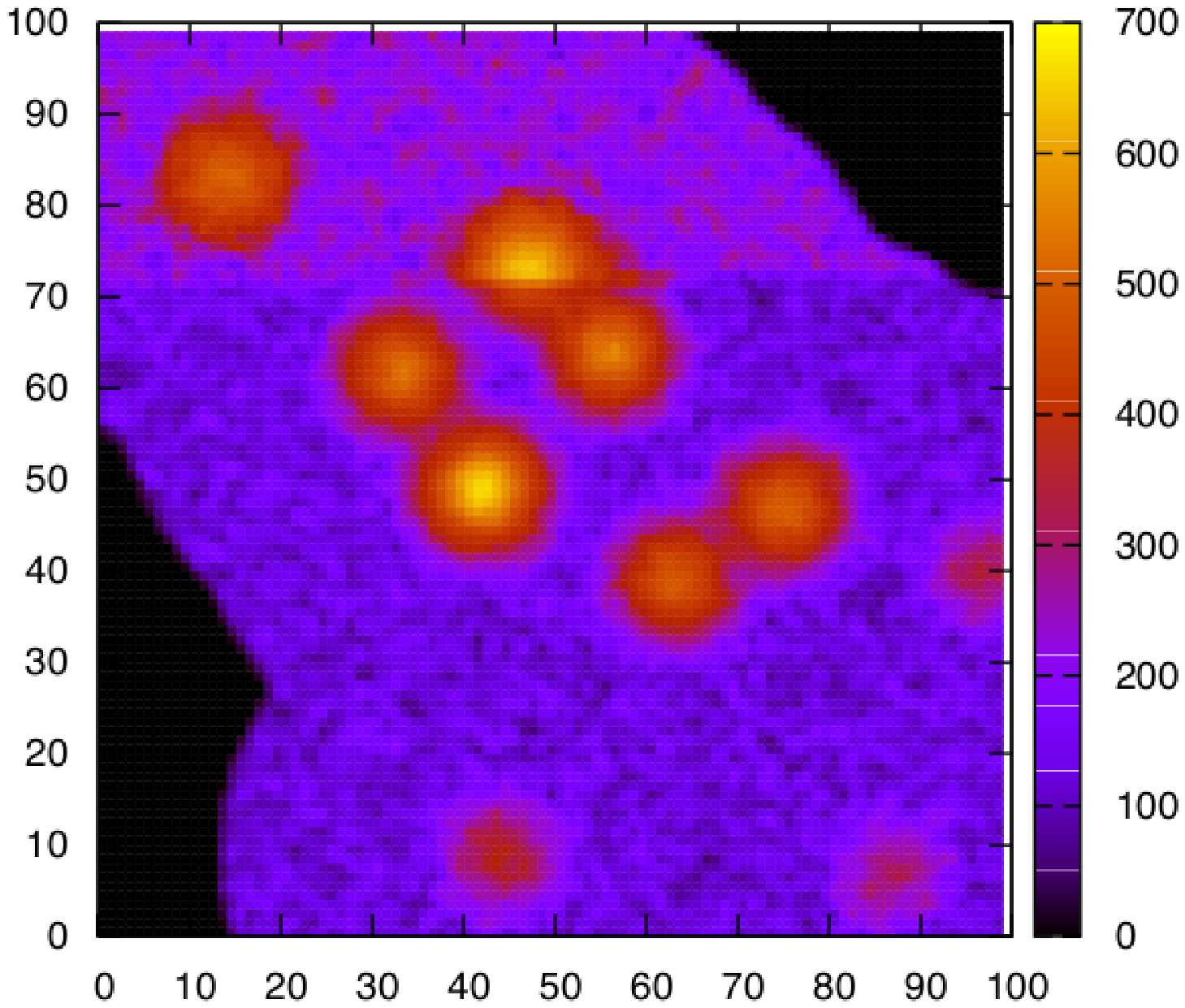} 
%\hspace{-0.3cm}
\includegraphics[width=5.6cm]{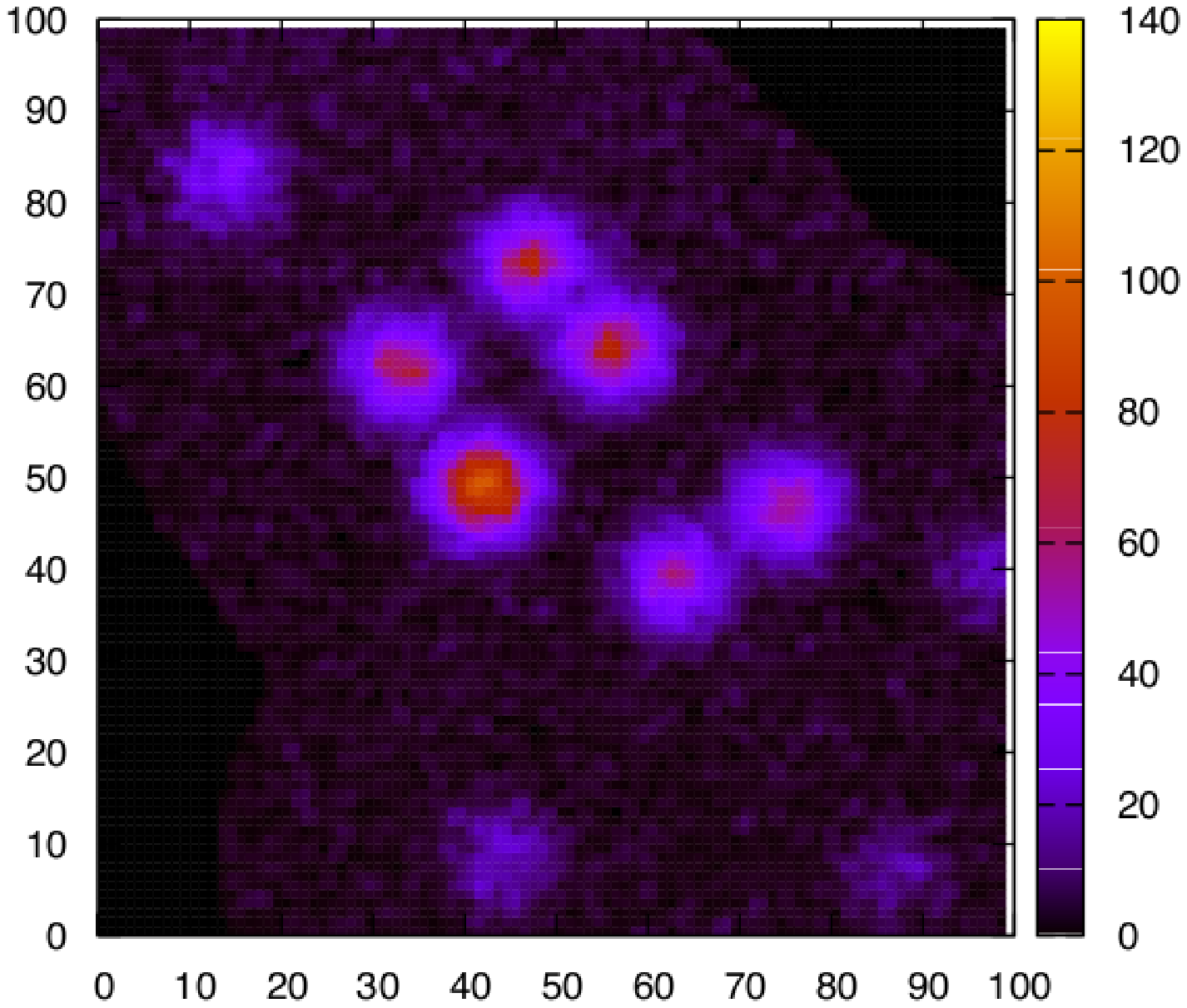} 
%\mbox{}\hspace{-0.2cm}
\includegraphics[width=5.6cm]{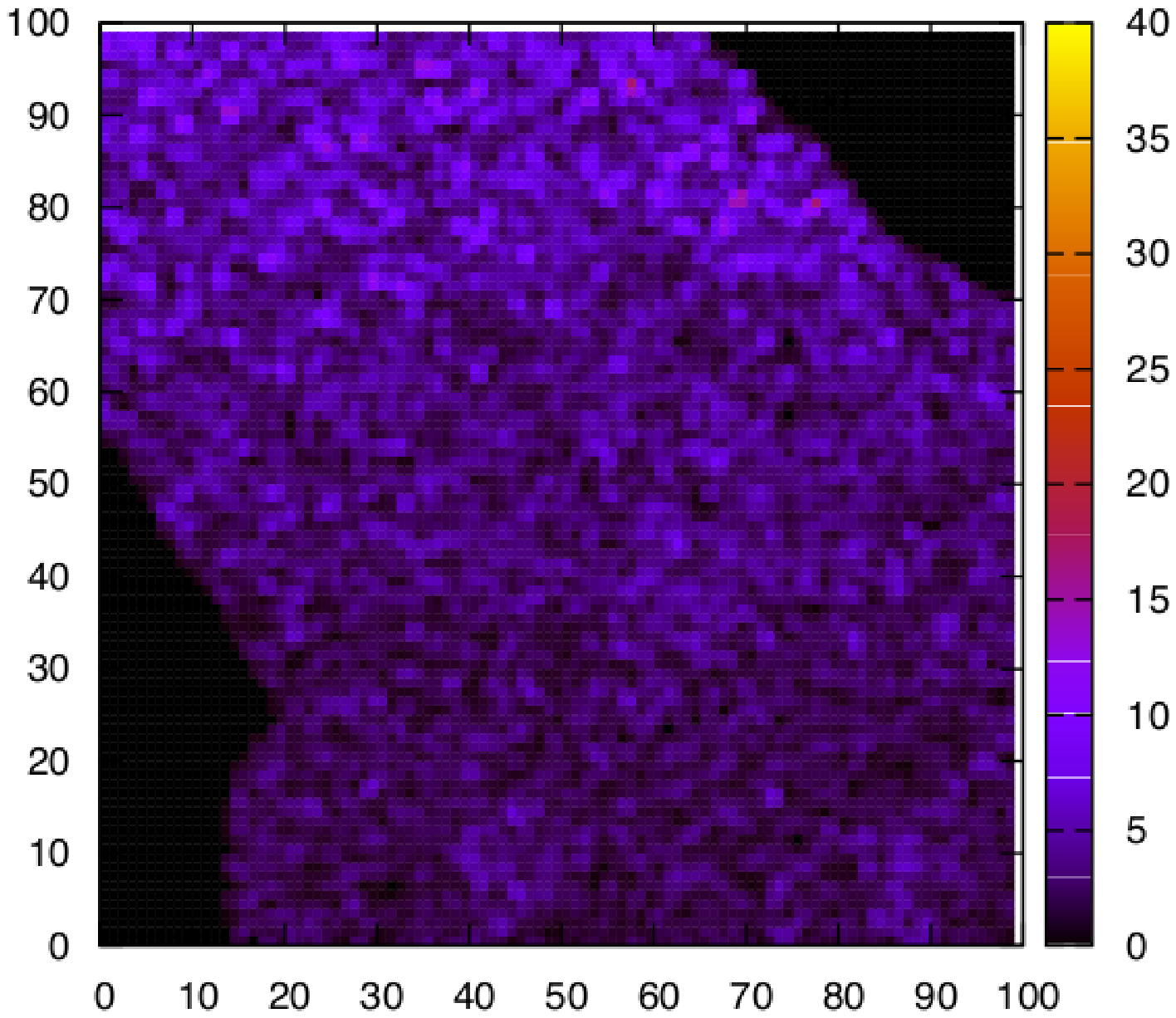} 
%\hspace{-0.3cm}
\includegraphics[width=5.6cm]{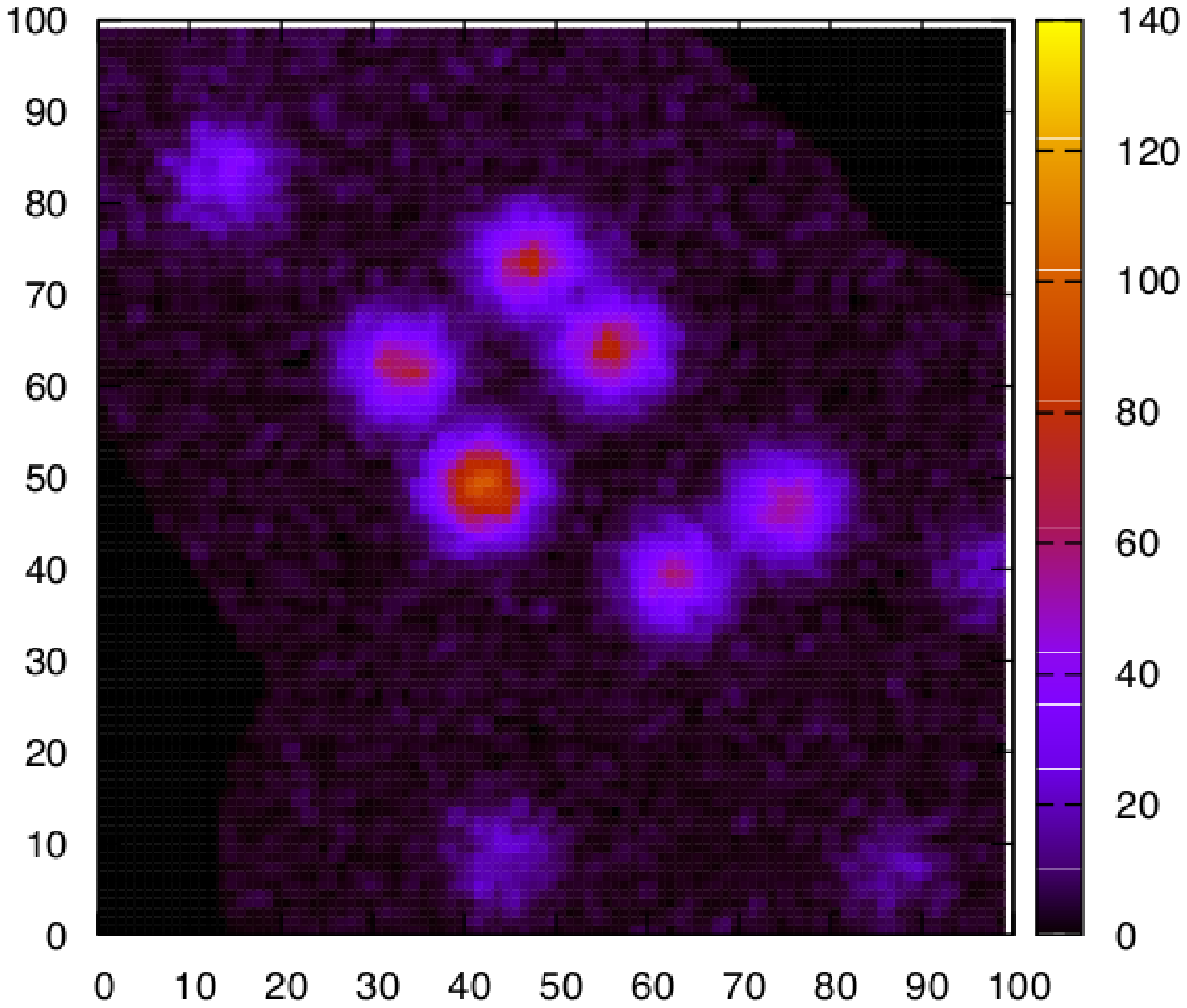} 
%%%%%%
\end{center}
\caption{\footnotesize 
The upper two panels show the spatial distribution 
of the number of visits consisting of 
the lowest ranking $k=0$ (left) and 
the highest ranking $k=9$ (right) agents. 
The corresponding 
spatial distributions of 
the number of the transaction approvals are shown 
in the lower panels. (COLOR ONLINE) 
}
\label{fig:agent_move}
\end{figure}

In order to investigate the explicit ranking dependence of 
the housing-search behavior of agents, 
in Fig. \ref{fig:agent_move} (the upper panels), 
we plot the spatial distribution 
of the number of visits for 
the lowest ranking $k=0$ (left) and 
the highest ranking $k=9$ (right) agents. 
We also plot the 
corresponding 
spatial distributions of 
the number of the transaction approvals 
in the lower two panels. 
From this figure, 
we confirm that 
the lowest ranking agents visit almost whole area of the city, 
whereas the highest ranking agents 
narrow down their visiting place for housing search. 
This result is naturally understood as follows. 
Although the lowest ranking agents visit some suitable places to 
live, they cannot afford to accept the offer price given by 
the sellers who are selling the flats at the place. 
As the result, 
such lowest ranking agents 
should wander from place to place to look for 
the place where the offer price is low enough for them to accept. 
That is a reason why the spatial distribution of 
visit for the lowest ranking agents distributes widely in the city. 
On the other hand, 
the highest ranking agents posses enough `willing to pay' $P_{9}$ and 
they could live any place they want. 
Therefore, their transactions are easily approved even at the centers of wards with 
relatively high intrinsic attractiveness $A^{0}(\bm{X})$. 
%%%%%%%%%%%%%%%%%
%%%%%%
\mbox{}

 As a non-trivial finding, 
 it should be noticed from Fig. \ref{fig:agent_move} that 
 in the northern part of the city (a part of {\it Kita} and {\it Higashi}-ku), 
 several small communities consisting of the lowest ranking 
 persons having their `willing to pay' $P_{0}$ emerge. In our modeling,  we do not use any `built-in' factor 
 to generate this sort of non-trivial structure. 
 This result might imply that 
 communities of poor persons could be emerged 
 in any city in any country even like Japan.  

Let us summarize our findings from simulation concerning 
ranking dependence of search-approvals by agents below. 
%%%
\begin{itemize}
\item
The lowest ranking agents ($k=0$) visit almost all of regions in city 
even though such places are highly `attractive places'. 
\item
The highest ranking agents ($k=9$) visit relatively high attractive places. 
The highest ranking agents are rich enough to afford to accept 
any offering price, namely,  $\#$ of contracts $\simeq$ $\#$ of visits. 
%%%
\item
The lowest ranking agents are swept away from relatively attractive regions and 
make several their own `communities' at low offering price locations in the city (the north-east area in Sapporo). 
\end{itemize}
%%%
%%%%%%%%%%%%%%%%%%%
%%%
%%%%%%%%%%%%%%%%%%%%%%%%%%%%%%%
\subsection{The rent distribution}
\begin{figure}[ht]
\begin{center}
%\mbox{}\hspace{-0.3cm}
%\includegraphics[width=5.5cm,bb=219 113 666 476]{FIGS/spatial_dist_without_office.pdf} 
\includegraphics[width=5.6cm]{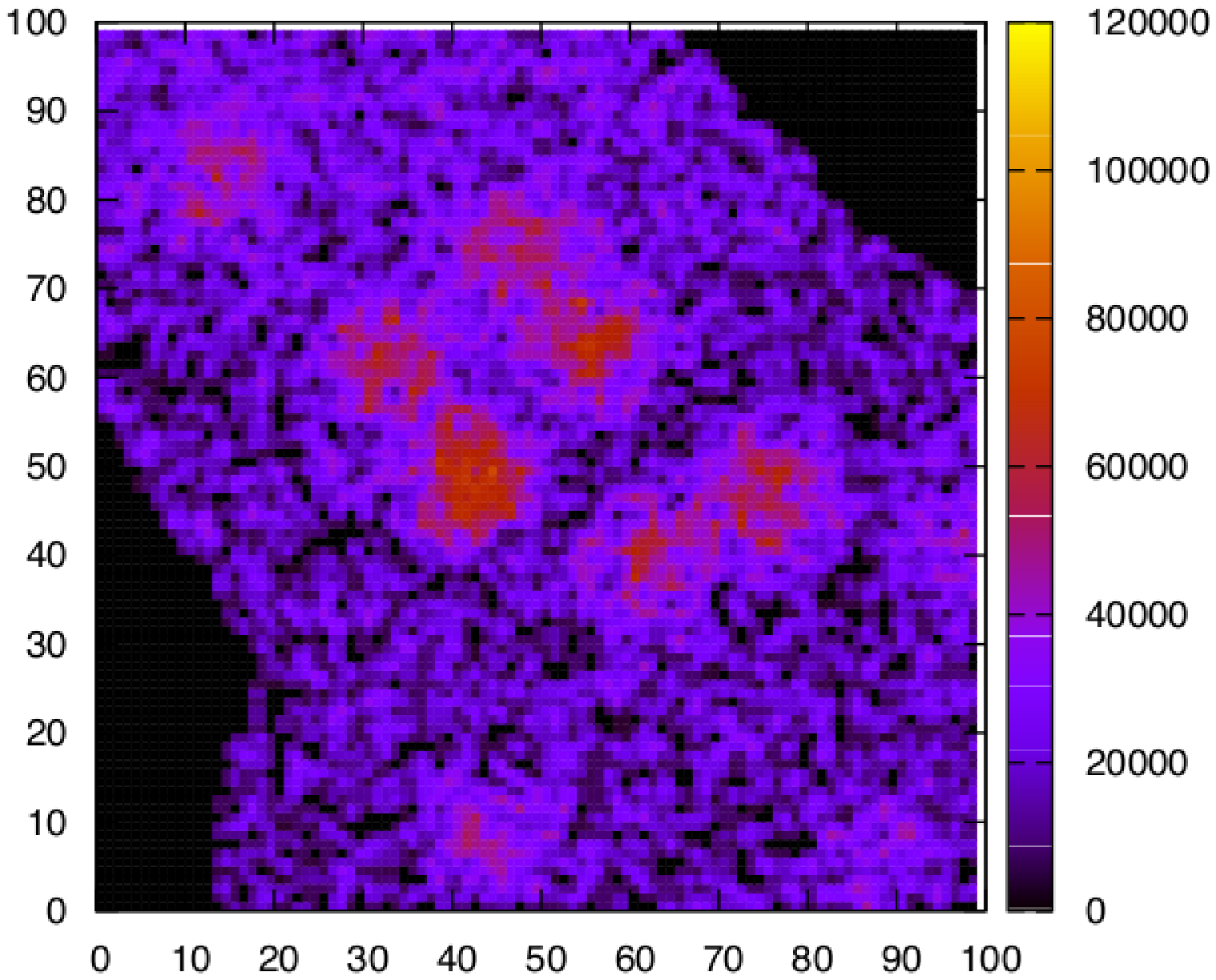} 
%\hspace{-0.2cm}
%\includegraphics[width=5.5cm,bb=223 115 662 475]{FIGS/spatial_dist_with_office.pdf}
\includegraphics[width=5.6cm]{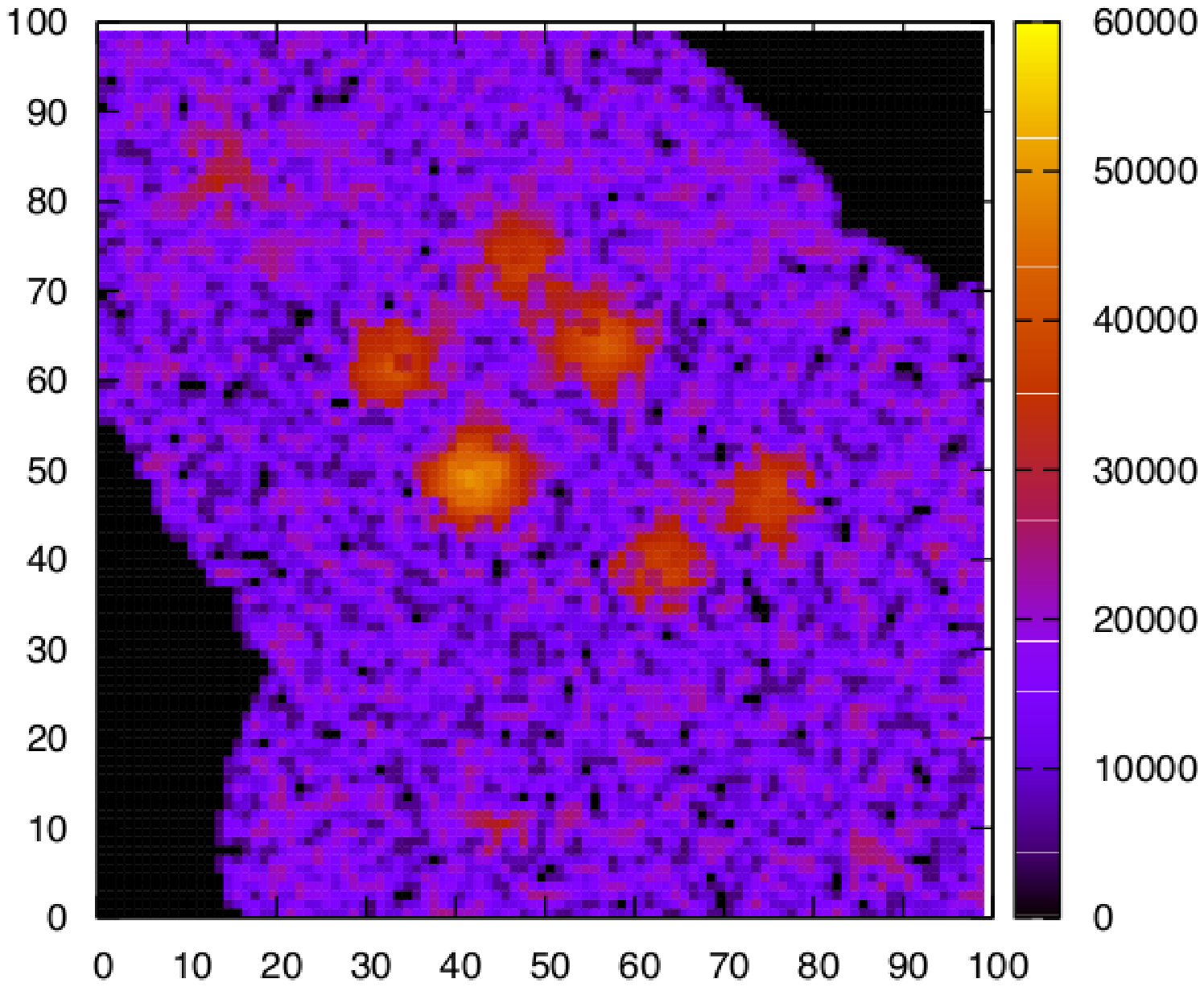}
\end{center}
%\mbox{}\vspace{-0.7cm}
\caption{\footnotesize 
The resulting 
spatial distributions of rent 
in city of Sapporo. 
In the left panel, 
we do not consider the effect of office 
locations on buyer's decision, whereas in the right panel, the effect is taken into account. (COLOR ONLINE) 
}
%\mbox{}\vspace{-0.5cm}
\label{fig:fg1}
\end{figure}

%%%%%%%%%%%%%%%%%%%%%%%%%%%%%%%%%%
In Fig. \ref{fig:fg1} (left), 
we plot the resulting spatial distribution of 
rent in city of Sapporo. 
From this figure we confirm that 
the spatial distribution is quantitatively similar to 
the empirical evidence. 
We also find that 
a complicated structure --- a sort of 
spatial anisotropy ---  emerges and 
it is completely different from the result by the Gauvin's model \cite{Nadal}. 
In particular, we should notice 
that relatively high rent regions 
around {\it Chuo}-ku appear. 
These regions are located near 
{\it Kita, Higashi, Nishi} and {\it Shiraishi}. 
%%%%%%%%%%%%%%%%%%%%%%
%%%%%%%%%%%%%%%%%%%%%%%
\mbox{}

To see the gap between our result and empirical evidence 
quantitatively, we show the simulated average rent and the counter-empirical evidence in Table \ref{tab:tb_order}. 
From this table, we find that 
the order of the top two wards simulated by our model, 
namely, {\it Chuo} and {\it Shiraishi} coincides with the empirical evidence, 
and moreover, the simulated rent itself is very close to 
the market price. 
However, concerning 
the order of the other wards, 
it is very hard for us to 
conclude that the model simulates the empirical data. 
Of course, 
the market price differences in those wards 
are very small and it is very difficult to simulate the correct ranking at present.  
Thus, the modification of our model to generate the correct ranking and 
to obtain the simulated rents which are much closer to the market prices should be addressed our future study. 
%%%%%%%%%%%%%%%%%%%%%%%%%%%%%%
\begin{table}[ht]
\begin{center}
\begin{tabular}{|l|r|}
\hline
ranking & market price (yen)\\ \hline
 {\it Chuo (Central)} & 54,200\\ \hline 
{\it Shiraishi} & 51,100 \\ \hline
{\it Nishi (West)} & 48,200 \\ \hline
{\it Kiyota} & 45,800 \\ \hline
{\it Teine} & 42,900 \\ \hline
{\it Kita (North)} & 41,700 \\ \hline
{\it Higashi (East)} & 40,100  \\ \hline
{\it Atsubetsu} & 39,700 \\ \hline
{\it Toyohira} & 39,600 \\ \hline
{\it Minami (South)} & 34,400 \\ 
\hline
\end{tabular}
%%%
\begin{tabular}{|l|r|}
\hline
ranking & simulated average rent (yen) \\ \hline
{\it Chuo (Central)} & 50,823 \\ \hline 
{\it Shiraishi} & 43,550 \\ \hline
{\it Higashi (East)} & 44,530 \\ \hline
{\it Kita (North)}  & 43,516 \\ \hline
{\it Nishi (West)} & 43,093 \\ \hline
{\it Toyohira} & 42,834 \\ \hline
{\it Minami (South)} & 39,909  \\ \hline
{\it Teine} & 39,775 \\ \hline
{\it Kiyota} & 37,041 \\ \hline
{\it Atsubetsu} & 36,711\\ 
\hline
\end{tabular}
\caption{\footnotesize 
The left  list shows the ranking (order) and 
the market prices. 
The simulated average rent for each ward and 
the ranking are shown in the right list. 
The unit of price is Japanese yen \cite{Homes}.}
\label{tab:tb_order}
\end{center}
%\mbox{}\vspace{-1.4cm}
\end{table}
%%%%%%%%%%%%
%%%%%%%%%%%%%%%%%%%%%%%%%%%%%%%%
\subsection{On the locations of offices}
\label{sec:remark}
%%%%%%%%%%%%%%%%%%%%%%%%%%%%%%%%%%
In the previous sections, 
we modified 
the intrinsic attractiveness 
so as to possess the multiple peaks 
at the corresponding 
locations of the ward offices by (\ref{eq:ours}). 
However, inhabitants must go to 
their office 
every weekday, 
and the location of office 
might give some impact 
on the decision making of each buyer in the city. 
For a lot of inhabitants in Sapporo city, 
their offices are located within the city, 
however, the locations are distributed. 
Hence, here we specify 
the ward in which his/her office is located 
by the label $m=1,\cdots,10$ and 
rewrite the intrinsic attractiveness (\ref{eq:ours}) as  
%%%%
\begin{eqnarray*}
A_{m}^{0}(\bm{X}) & = &  
\sum_{l \neq m}^{10}
\frac{\delta_{l}}{\sqrt{2\pi} R_{l}}
{\exp}
\left[
-\frac{
\{
(x-x_{B_{l}})^{2}+
(y-y_{B_{l}})^{2}
\}
}{2R_{l}^{2}}
\right] \nonumber \\
\mbox{} & + & 
\frac{(\delta_{m} + \eta)}{\sqrt{2\pi} R_{m}}
{\exp}
\left[
-\frac{
\{
(x-x_{B_{m}})^{2}+
(y-y_{B_{m}})^{2}
\}
}{2R_{m}^{2}}
\right]. 
\label{eq:ours2}
\end{eqnarray*}
%%%%%
Namely, 
for the buyer 
who has his/her office within 
the ward $m$, 
the ward $m$ might be a `special region' for him/her and 
the local peak appearing in the intrinsic attractiveness 
is corrected by $\eta$. 
If he/she seeks for the housing close to his/her house (because 
the commuting cost is high if the office is far from his/her house), 
the correction $\eta$ takes a positive value. 
On the other hand, 
if the buyer wants to live the place located far from the office 
for some reasons (for instance, 
some people want to vary the pace of their  life), 
the correction $\eta$ should be negative.  
To take into account these naive assumptions, 
we might choose 
$\eta$ as a snapshot from 
a Gaussian with mean zero and 
the variance $\sigma^{2} (< \delta_{m})$. 
From this type of 
corrections, 
the buyer, in particular,  the buyer 
of the highest ranking ($k=K-1$) 
might feel some `frustration' to 
make their decision, 
which is better location for them between 
{\it Chuo}-ku as the most attractive ward and 
the ward $m$ where his/her office is located under 
the condition $\delta_{1} \simeq \delta_{m}+\eta$. 
For the set of weights 
$\delta_{m}, \,m=1,\cdots,10$, 
we take into account the 
number of offices in each ward, 
that is, 
{\it  Chuo} (23,506), 
{\it Kita} (8,384), 
{\it Higashi} (8,396), 
{\it Shiraishi} (7,444), 
{\it Toyohira} (6.652), 
{\it Minami}  (3,418), 
{\it Nishi} (6,599), 
{\it Atsubetsu} (2,633), 
{\it Teine} (3,259), 
{\it Kiyota} (2,546). 
Then, we choose 
each $\delta_{m}$ by 
dividing 
each number by the maximum of {\it Chuo}-ku 
as  
{\it Chuo} ($\delta_{1} \propto 1.00$), 
{\it Higashi} ($\delta_{2} \propto 0.36$), 
{\it Nishi} ($\delta_{3} \propto 0.28$), 
{\it Minami} ($\delta_{4} \propto 0.15$), 
{\it Kita} ($\delta_{5} \propto 0.36$), 
{\it Toyohira} ($\delta_{6} \propto 0.28$), 
{\it Shiraishi} ($\delta_{7} \propto 0.32$), 
{\it Atsubetsu} ($\delta_{8} \propto 0.11$), 
{\it Teine} ($\delta_{9} \propto 0.14$), 
{\it Kiyota} ($\delta_{10} \propto 0.11$). 
For the bias parameter $\eta$, 
we pick up the value randomly from the range: 
%%%%
%%%
\begin{equation}
|\eta| < \delta_{m}, 
\label{eq:work2}
\end{equation}
%%%%%
instead of the Gaussian. 

The resulting 
spatial distribution is shown in the right panel of 
Fig. \ref{fig:fg1}. 
From this panel, we are clearly confirmed that 
the spatial structure of rents distributes more widely in whole city than 
that without taking into account the office location (see the left panel in Fig. \ref{fig:fg1} for comparison). 
We should notice that 
the range of simulated rent in the city is remarkably 
reduced from $[0,120,000]$ to $[0,60,000]$ 
due to the diversification of values to consider the location of their housing.
%%%%%%%%%%%%%%%%%%%%%%%%%%%
\subsection{On the effective time-scale of update rule}
%%%%%%%%%%%%%%%%%%%%%%%%%%% 
Until now, we did not make a mention of time scale 
in the spatio-temporal update rule (\ref{eq:update}) in  the attractiveness $A_{k}(\bm{X},t)$. 
However, 
it might be important to for us consider 
how long the time in our model system (artificial society) 
goes on 
for the minimum time step $t \to t+1$, 
especially when we evaluate the 
necessary period of time to complete the accumulation of 
community after new-landmarks or shopping mall come out. 
To decide the effective time-scale for $t \to t+1$, 
we use the information about the number of persons moving-into city of Sapporo 
through the year. 
Let us define the number from the empirical data by $C$. 
Then, we should remember that 
in our simulation, we assumed that 
in each time step ($t \to t +1$), 
$\Gamma$ newcomers visit the city.
Hence, 
the actual time $\tau$ for the minimum time step $t \to t+1$ in 
our artificial society is effectively given by 
%%%%
\begin{equation}
\tau =
365 \times \frac{\Gamma}{C}\,\,[\mbox{days}]. 
\label{eq:def_time}
\end{equation}
%%%
Therefore, 
by using our original set-up $\Gamma \equiv L^{2}/K=(100\times 100)/10=1000$ and  
by making use of 
the data listed in Table \ref{tab:population}, 
we obtain $C=63021$, and 
substituting the value into (\ref{eq:def_time}), 
we finally have 
$\tau = (1000 \times 365)/63021=5.79$ [days] for $t \to t +1$. 
This means that 
approximately 579 days have passed when 
we repeat the spatio-temporal update rule (\ref{eq:update}) by $T=100$ times. 
This information might be essential when we predict the future housing market, let us say,  
after constructing the {\it Hokkaido Shinkansen} (a rapid express in Japan) railway station, 
related landmarks and derivative shopping mall. 
%%%%%%%%%%%%%%%%%%%%%%%%%%%%%%%%%%%
\section{Summary and discussion}
\label{sec:summary}
%%%%%%%%%%%%%%%%%%%%%%%%%%%%%%%%%%%%
In this paper, we modified 
the Gauvin's model \cite{Nadal} to 
include the city having multiple centers 
such as the city designated by ordinance 
by correcting the intrinsic attractiveness $A^{0}(\bm{X})$. 
As an example for such cities, 
we selected our home town Sapporo and 
attempted to simulate the spacial distribution 
of averaged rent. 
We found that our model can explain the empirical evidence 
qualitatively. 
Especially, we found that the lowest ranking agents (from the viewpoint of the lowest `willing to pay') are 
swept away from relatively attractive regions and make several their own `communities' at  
low offering price locations in the city. 

However, we should mention 
that we omitted an important aspect in our modeling. 
Namely, the spatial resolution of working space and probabilistic search 
by buyer taking into account their office location. 
In following, we will make remarks on those two issues. 
%%%%%%%%%%%%%%%%%%%%%%%%%%%%%%%%%
\subsection{ The `quasi-one-dimensional' model}
\begin{figure}[ht]
\begin{center}
\includegraphics[width=10cm]{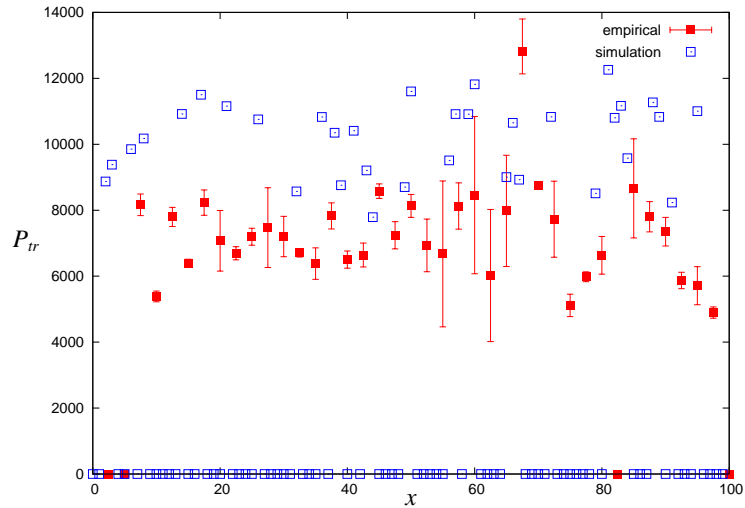} 
%\mbox{}\vspace{-0.7cm}
\end{center}
\caption{\footnotesize 
The resulting rent distribution along the {\it Tozai}-subway line 
obtained by the quasi-one-dimensional model. 
The correspondence between $x$-axis and 
the location of each subway station is explicitly given as follows: 
{\it Bus-center mae} ($x=86$),
{\it $\bar{O}$dori} ($x=76$),
{\it Nishi 11-Choume} ($x=54$),
{\it Nishi 18-Choume} ($x=36$),
{\it Maruyama Kouen} 
($x=22$), and 
{\it Nishi 28-Choume} ($x=11$). 
 }
\label{fig:fg_dim}
\end{figure}

%%%%%%%%%%%%%%%%%%%%%%%%%%%%%%%%%%
The problem of spatially low resolution of working space 
might be overcame when we model the housing market  focusing on 
{\it Chuo}-ku instead of whole (urban) part of Sapporo. 
In the modeling, we restrict ourselves to the `quasi-one-dimensional' 
working space and 
this approach enables us to 
compare the result with the corresponding empirical data. 
Although it is still at the preliminary level, 
we show 
the resulting rent distribution 
along the {\it Tozai}-subway line 
which is running across the center of Sapporo ({\it Chuo}-ku) 
from the west to the east as in Fig. \ref{fig:fg_dim}. 
Extensive numerical studies  
accompanying with 
collecting data with higher resolution 
are needed to proceed the present study and 
it should be addressed as our future work.  
%%%%%%%%%%%%%%%%
%%%%%%%%%%%%%%%%
%%%%%%%%%%%%%%%%%%%%%%%%%%%%%%%%%%%%%%%%%%%%
\subsection{Probabilistic search depending on the location of office}
%%%%%%%%%%%%%%%%%%%%%%%%%%%%%%%%%%%%%%%%%%%%
Some of buyers might search 
their housing locations by taking into account the place of their office. 
Here we consider the attractiveness $B_{k,i}(\bm{X},t)$ of office location $\bm{X}$ for $i$ ranking $k$ at time $t$. 
Namely, 
we should remember that 
the attractiveness for the place to live is updated as 
%%%%%%%%%%%%%%%%%%
\begin{equation}
A_{k}(\bm{X},t+1) = 
A_{k}(\bm{X},t) + \omega 
(A^{0}(\bm{X})-A_{k}(\bm{X},t))
+\epsilon \sum_{k^{'} \geq k}v_{k}^{'}(\bm{X},t) 
\end{equation}
%%%
depending on 
the intrinsic attractiveness 
of place to live, 
whereas the attractiveness 
of the place $\bm{X}$ for agent $i$ of ranking $k$ 
who takes into account the location of their 
office place $\bm{Y}_{k,i}$ is also defined accordingly 
and it is governed by 
%%%%
\begin{equation}
B_{k,i}(\bm{X},t+1)  =  
B_{k,i}(\bm{X},t) + 
\overline{\omega}
(B_{k,i}^{0}(|\bm{X}-\bm{Y}_{k,i}|)-
B_{k,i}(\bm{X},t)
)+\overline{\epsilon}\sum_{k^{'} \geq k}
v_{k^{'}}(\bm{X},t)
\label{eq:updateB}
\end{equation}
%%%%
where $B_{k,i}^{0}(|\bm{X}-\bm{Y}_{k,i}|)$ is `intrinsic attractiveness'  of the location $\bm{X}$ for the agent 
$i$ whose office is  located at $\bm{Y}_{k,i}$ 
and it is given explicitly by 
%%%
\begin{equation}
B_{k,i}^{0}(|\bm{X}-\bm{Y}_{k,i}|) = 
\frac{1}{\sqrt{2\pi}Q}
{\exp}
\left[
-\frac{(\bm{X}-\bm{Y}_{k,i})^{2}}
{2\pi Q^{2}}
\right], 
\end{equation}
%%%%%%%%%%
and $\overline{\omega}, \overline{\epsilon}$ are 
parameters to be calibrated using the empirical data sets. 

Then, agent $i$ looks for the candidates $\bm{X}_{A},\bm{X}_{B}$ to live according to the 
following probabilities 
%%%%
\begin{eqnarray}
\pi_{k}^{(A)}(\bm{X},t) & = &  
\frac{1-{\exp}(-\lambda A_{k}(\bm{X},t))}
{\sum_{\bm{X}^{'} \in \Omega}
\{1-{\exp}(-\lambda A_{k}(\bm{X}^{'},t))\}} 
\label{eq:pi_A} \\
%%%%
\pi_{k,i}^{(B)}(\bm{X},t) & = &  
\frac{1-{\exp}(-\lambda B_{k,i}(\bm{X},t))}
{\sum_{\bm{X}^{'} \in \Omega}
\{1-{\exp}(-\lambda B_{k,i}(\bm{X}^{'},t))\}}
\label{eq:pi_B}
\end{eqnarray}
%%%%%
%%%
If the both $\bm{X}_{A},\bm{X}_{B}$ are approved, transaction price for each place is given by 
\begin{eqnarray}
P_{\rm tr}^{(A)} & = & 
(1-\beta) P_{k^{'}}(\bm{X}_{A}) + 
\beta P_{k} \\
%%%
P_{\rm tr}^{(B)} & = & 
(1-\beta) P_{k^{'}}(\bm{X}_{B}) + 
\beta P_{k} 
\end{eqnarray}
%%%
\begin{figure}
\begin{center}
\includegraphics[width=8cm]{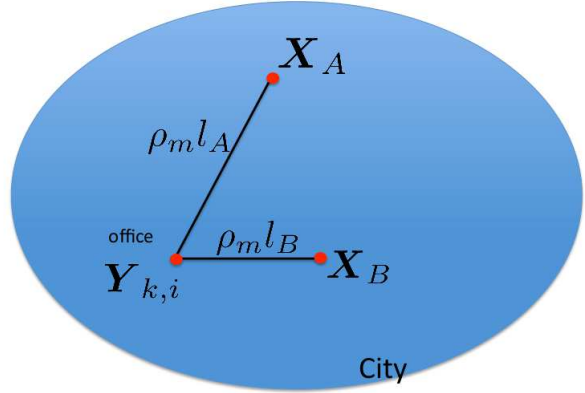}
\end{center}
\caption{\footnotesize 
Each buyer decides 
the final location 
$\bm{X}_{F}={\arg\min}\{P_{\rm tr}^{(A)} +\rho_{m} l_{A},
, P_{\rm tr}^{(B)} + 
\rho_{m}l_{B}
\}, F=\{A,B\}$. 
}
\label{fig:fg_prob_B}
\end{figure}
%%%%%%%
The finial decision $\bm{X}_{F}$ is 
\begin{equation}
\bm{X}_{F}={\arg\min}\{P_{\rm tr}^{(A)} +\rho_{m} l_{A},
, P_{\rm tr}^{(B)} + 
\rho_{m}l_{B}
\}, \, F=\{A,B\}
\end{equation}
%%%
where $\rho_{m} l_{A},\rho_{m} l_{B}$ are 
travel costs between $\bm{X}_{A}, \bm{X}_{B}$ and office 
(see Fig. \ref{fig:fg_prob_B}). 

Therefore, 
the agents might prefer relatively closer place to the office 
to the attractive place to live 
when the distance between the attractive place to live and 
the office is too far for agents to manage the cost by the commuting allowance. 
%%%
%%%%%%%%%%%%%%%%%%%%%%%%%%%%%%
%\section*{Acknowledgments}
%%%%%%%%%%%%%%%%%%%%%%%%%%%%%%%%%%%%%%%%%
\begin{acknowledgement}
One of the authors (JI) 
thanks  Jean-Pierre Nadal in $\acute{\rm E}$cole Normale Sup$\acute{\rm e}$rieure 
for fruitful discussion on this topic and useful comments on our preliminary results  
at the international conference 
{\it Econophysics-Kolkata VII}. 
The discussion with Takayuki MIzuno, Takaaki Onishi, 
and Tsutomu Watanabe was very helpful to prepare this manuscript. 
This work was financially supported by 
Grant-in-Aid for Scientific Research (C) 
of Japan Society for 
the Promotion of Science(JSPS) No. 22500195, 
Grant-in-Aid for Scientific Research (B) No. 26282089, 
and Grant-in-Aid for Scientific Research on Innovative Area No. 2512001313. 
Finally, we would like to acknowledge the organizers of {\it Econophys-Kolkata VIII} for their hospitality during the conference, 
in particular, Frederic Abergel, Hideaki Aoyama, Anirban Chakraborti, 
Asim Ghosh and Bikas K. Chakrabarti. 
\end{acknowledgement}

%%%%%%%%%%%%%

\end{document}